\documentclass{JHEP3} 
\usepackage{epsfig} 
\usepackage{graphicx}

\usepackage{bm}
\usepackage{amssymb}
\usepackage{amsmath}

\def\fun#1#2{\lower3.6pt\vbox{\baselineskip0pt\lineskip.9pt
  \ialign{$\mathsurround=0pt#1\hfil##\hfil$\crcr#2\crcr\sim\crcr}}}

\def\fun#1#2{\lower3.6pt\vbox{\baselineskip0pt\lineskip.9pt
\ialign{$\mathsurround=0pt#1\hfil##\hfil$\crcr#2\crcr\sim\crcr}}}

\def\mpl{m_{\rm Pl}}
\newcommand{\MUNCH}[1]{\relax}

\title{Slow Roll Reconstruction: Constraints on Inflation from the 3 Year WMAP 
Dataset}

\author{Hiranya V. Peiris\thanks{Hubble Fellow}  \\
Kavli Institute for Cosmological Physics and Enrico Fermi Institute, \\University of Chicago, Chicago IL 60637, USA. \\
\email{hiranya@cfcp.uchicago.edu}}

 \author{Richard Easther \\
Department of Physics, Yale University, New Haven CT 06520, USA \\
\email{richard.easther@yale.edu}}

\abstract{We study the constraints on the inflationary parameter space derived from the 3 year WMAP dataset using ``slow roll  reconstruction'', using the SDSS galaxy power spectrum to gain further leverage where appropriate.  This approach inserts the inflationary slow roll  parameters directly into a Monte Carlo Markov chain estimate of the  cosmological parameters, and uses the inflationary flow hierarchy to compute the parameters' scale-dependence.   We work with the first three parameters ($\epsilon$, $\eta$ and $\xi$) and pay close attention to the possibility that the 3 year WMAP  dataset contains evidence for a ``running'' spectral index, which is dominated by the $\xi$ term.  Mirroring the WMAP team's analysis we find that the permitted distribution of $\xi$ is broad, and centered away from zero. However, when we require that inflationary parameters yield at least 30 additional e-folds of inflation after the largest observable scales leave the horizon, the bounds on $\xi$ tighten dramatically.  We make use of the absence of an explicit pivot scale in the slow roll reconstruction formalism to determine the dependence of the computed parameter distributions on the pivot. We show that the choice of pivot has a significant effect on the inferred constraints on the inflationary variables, and the spectral index and running derived from them. Finally, we argue that the next round of cosmological data can be expected to place very stringent constraints on the region of parameter space open to single field models of slow roll inflation.
}
\preprint{}
\begin{document}

\section{Introduction}

The primordial perturbation spectrum poses two distinct problems  for cosmology. Observationally, we wish to extract the  spectrum's amplitude and scale dependence from data, while theoretically we seek to understand the origin of  the perturbations.   After the release of the 3-year WMAP dataset \cite{Hinshaw:2006ia,Page:2006hz,Spergel:2006hy} there is considerable optimism that we will soon  measure the departure from scale  invariance in the spectrum at significance levels greater than   $3 \sigma$.\footnote{WMAP is an acronym for the {\sl Wilkinson Microwave Anisotropy Probe}. Here WMAPI refers to the first year's data from the satellite, and WMAPII refers to the second WMAP data release, which is drawn from the first three years' data.}     On the theoretical front, inflation \cite{Guth:1980zm,Albrecht:1982wi,Linde:1981mu}  remains the leading theoretical paradigm for understanding  the very early universe and  the origin of the perturbations.  In general terms, the latest round of cosmological data has confirmed the broad predictions of inflation -- the universe is very close to being flat, it has a roughly scale-invariant perturbation spectrum and $\langle TE \rangle$, the  cross correlation between the  E mode of the polarization and the temperature, has the form predicted by inflation. 

Beyond this rosy picture, however, lies a number of qualifications. Firstly, the observed form of  $\langle TE \rangle$ is not a unique prediction of inflation, but is also a generic feature  of models where the perturbations are laid down during a collapsing phase prior to a bounce, such as the ekpyrotic scenario \cite{Khoury:2001wf}.\footnote{The $\langle TE \rangle$ signal  is the same for any mechanism that induces adiabatic perturbations that are correlated on scales much larger than the post-recombination horizon volume.  ``Bouncing'' models do this by laying down perturbations during the contracting phase, when the comoving horizon size is shrinking. }   Secondly, inflation can be implemented in a vast number of different ways, and each mechanism produces a potentially distinctive perturbation spectrum. Consequently, the inflationary paradigm does not make an unambiguous prediction for the detailed form of the perturbation spectrum.  Some authors have posited that inflation has a ``default'' perturbation spectrum \cite{Boyle:2005ug}.  The underlying arguments leading to this position rely on simplicity and naturalness, rather than a watertight ``no go'' theorem.  Consequently, inflation itself will not be ruled out if the perturbation spectrum turns out to differ from this simple expectation, but it is certainly true that many well-known models   would be excluded.     

Looking at data, it is always tempting to focus on possible anomalies -- parameters that differ noticeably from their ``expected'' values, but for which the discrepancy falls well short of a $3\sigma$ detection. Since the release of the original WMAPI dataset \cite{Bennett:2003bz,Spergel:2003cb,Peiris:2003ff}, three aspects of the power spectrum have received considerable attention: the ``low''  values of $C_\ell$ at small $\ell$, the glitches around the first Doppler peak and at $\ell \sim 40$, and the apparent evidence for a running, or scale-dependent, spectral index that arose when the WMAPI data was combined with other probes of the power spectrum at shorter scales.   The status of all of these anomalies has evolved since the WMAPI release. The quadrupole is still $\sim 2 \sigma$ lower than the LCDM best-fit value, but the octupole has moved closer to the LCDM value, thanks to an optimal extraction of the spectrum.  The glitch near the first Doppler peak is not present in the WMAPII dataset with its smaller errorbars, while the $\ell \sim 40$ discrepancy is still seen \cite{Cline:2006db,Covi:2006ci}.  Finally, the WMAPII dataset contains a preference for a running spectral index  when viewed alone, in combination with other CMB data, and/or large scale structure information such as 2dFGRS \cite{Cole:2005sx} and the Sloan Digital Sky Survey \cite{Tegmark:2003uf}.  However, a reanalysis by Seljak {\em et al.}, which includes SN1a data and information from the Lyman-$\alpha$ forest, puts a tight constraint on any running \cite{Seljak:2006bg} --- this result seems to hinge on the inclusion of the Lyman-$\alpha$ data. Further analysis of a possible running index is provided by \cite{Feng:2006ui}.
  
In this paper, we focus on the running of the spectral index, and the constraints it places on the inflationary parameter space.  Conventionally, one specifies the spectrum at a pivot point in terms of the spectral index $n_s$ and the running $\frac{dn_s}{d\ln{k}}$.  Inflation predicts $n_s$ and $\frac{dn_s}{d\ln{k}}$ in terms of the slow roll parameters, which are functions of the potential and its derivatives.   In \cite{Easther:2006tv}  we showed that if the running is as large as the central value extracted from the WMAPII dataset,  no {\em simple\/} model of inflation can simultaneously provide both this running and a sufficient amount of inflation.\footnote{Which we take to be a minimum of 30 e-foldings, and by  ``simple'' we mean a single inflaton, minimally coupled to gravity, whose potential can be well specified in terms of its first three derivatives.}  Currently, there are simple models of inflation within the errorbars, but if the constraints from the data continue to tighten around the median value of the running, they will be ruled out. We arrived at this result by considering the Hubble Slow Roll hierarchy, otherwise known as the inflationary flow equations \cite{kinney:2002,easther/kinney:2003,liddle:2003,Peiris:2006ug}.  These codify the scale dependence of the slow roll parameters: given the values of the slow roll parameters at  a single moment during the inflationary era, we can compute their values at all other times.  A further virtue of this approach is that when the infinite slow roll hierarchy is truncated at finite order, the truncation is preserved by the evolution equations. Using the slow roll hierarchy means that we effectively dispense with the pivot point, other than when comparing our results to treatments using the standard $n_s$ and $\frac{dn_s}{d\ln{k}}$ parameterization of the spectrum.

As we  showed in \cite{Peiris:2006ug}, we can include the slow roll parameters $\{\epsilon,\eta,\xi,\cdots\}$ in the cosmological parameter estimation process, and entirely avoid the use of the spectral index  and its running. Further, the slow roll hierarchy can be solved to produce the underlying inflationary potential, leading us to dub this approach ``slow roll reconstruction.''\footnote{The overall amplitude of the spectrum  is fixed by an arbitrary constant of integration, which can only be measured directly if the tensor spectrum is observed.}    Of course, this presupposes that the perturbations were produced by inflation, so slow roll reconstruction does not replace fits to $n_s$ and $\frac{dn_s}{d\ln{k}}$, which are purely empirical characterizations of the spectrum and contains no strong theoretical priors.\footnote{The index and running are effectively the first two terms in a  Taylor series expansion, so their use does make the tacit assumptions that the spectrum is a smooth, differentiable function of the comoving wavenumber $k$.} However, if we do adopt an inflationary prior, fitting to the slow roll parameters directly and using the flow equations to obtain their scale dependence ensures that we gain the maximum possible leverage from imposing this constraint.  It is well known that the scalar and tensor perturbations obey a consistency condition at lowest order in slow roll,\footnote{In practice, almost all estimates of the cosmological parameters assume that the amplitude and slope of the tensor spectrum have the correlation predicted by minimal models of inflation. Since the tensor spectrum is merely bounded from above,  rather than observed directly,  this constraint makes a Monte Carlo Markov Chain  analysis more tractable without excluding viable models.} but there are further, more subtle correlations at second order and beyond which tie together the running and  the  scalar spectral index \cite{Cortes:2006ap}.  By using slow roll reconstruction we automatically include these correlations in our analysis.  Moreover, this removes a potential source of ambiguity inherent in the introduction of a scale dependent spectral index -- which is how to best parametrize the scale dependence?  While $\frac{dn_s}{d\ln{k}}$ is arguably the simplest measure of the scale dependence, it is by no means unique. In particular, by adopting it we are implicitly assuming that while $n_s$ varies with $k$, $\frac{dn_s}{d\ln{k}}$ does not.  By specifying the spectrum in terms of the slow roll parameters we ensure our spectrum has the scale dependence predicted by inflation, and the primary limitation is the highest order  slow roll parameter we include in our fit. While this is obviously a choice that will be guided by the quality of the data, it means that we can describe the scale dependence in a relatively unambiguous way.\footnote{This issue mirrors the concern when the dark energy equation of state is allowed to vary with time, as there is no unique way of specifying that time dependence \cite{Corasaniti:2002vg,Linder:2005ne}.}   Finally, since we obtain the overall form of the potential as a by-product of this analysis, we can include constraints based on the total number of e-folds of inflation.  

The closest analogue to slow roll reconstruction is the approach of Leach and collaborators using the Horizon Flow Functions \cite{Leach:2002ar,Leach:2002dw,Leach:2003us}, who compute the spectral index and running as functions of the slow roll parameters. As in slow roll reconstruction, one has an explicit inflationary prior, and the slow roll parameters appear directly in the parameter set.  The advantage of this approach is that it can be implemented more simply in existing codes, since it translates between the slow roll formalism and  $n_s$ and $\frac{dn_s}{d\ln{k}}$, whereas in order to implement slow roll reconstruction we have to numerically solve the differential equation that maps the field evolution into the comoving wavenumber $k$.\footnote{In practice this is not computationally expensive, especially when compared to the numerical evaluation of the CMB spectrum and likelihood codes.}
However, slow roll reconstruction can be more easily extended to include higher order terms in slow roll, and copes naturally with heterogeneous datasets that span a wide range of wavelengths -- this is not impossible in the formalism of \cite{Leach:2002ar,Leach:2002dw,Leach:2003us}, but would require  more algebraic labor. Secondly, slow roll reconstruction can naturally include constraints based on the total duration of inflation. Finally, and perhaps most importantly, because \cite{Leach:2002ar,Leach:2002dw,Leach:2003us} use $n_s$ and $\frac{dn_s}{d\ln{k}}$, they are effectively Taylor expanding the power spectrum round a fixed pivot scale, $k_0$.  Conversely,  we can easily post-process our chains to present parameter constraints  at an arbitrary pivot.   In \cite{Finelli:2006fi}, the formalism of  \cite{Leach:2002ar,Leach:2002dw,Leach:2003us}  is applied to the WMAPII dataset to produce constraints on the slow roll parameters, and this work is directly comparable to the results we present below.   Slow roll reconstruction also draws on the stochastic Monte Carlo reconstruction algorithm, described in \cite{easther/kinney:2003}.  This procedure does not yield unambiguous estimates of the slow roll variables, but it permits estimates of the impact of higher order terms  in the slow roll expansion  \cite{Peiris:2006ug} and has been applied to cosmological data in \cite{Peiris:2003ff,Kinney:2003uw,Malquarti:2003ia,Kinney:2006qm}; a similar approach can also be found in \cite{Hansen:2001eu,Caprini:2002jy}.
An alternative approach of computing exact power spectra for specific potentials has been applied to the WMAPII data by \cite{Martin:2006rs}.

Our specific goal here is to extend our analysis in \cite{Peiris:2006ug} to include the third  slow-roll parameter, $\xi$, in a Monte Carlo Markov Chain  [MCMC] analysis.  This parameter dominates  the running of the spectrum, so it must be incorporated in any discussion of a running spectral index, especially in the light of the hints in the WMAPII data that the spectrum has a non-trivial scale-dependence.  We will also include constraints based on $N$, the number of e-folds of inflation remaining after the fiducial scale has left the horizon.       We consider three different sets of chains:  WMAPII + $N>30$, WMAPII + SDSS + $N>30$, and WMAPII + SDSS with no $N$ constraint. While we put a cut on $N$, we do not take the further step of correlating $N$ with the inflationary energy scale.  A low $N$ generally implies that inflation occurs well below the GUT scale but making this statement precise requires introducing assumptions about the physics of reheating, which we wish to avoid.\footnote{For instance, if the post-inflationary  universe has an epoch in which  $a(t) \sim t$ -- as might happen if the end of inflation is followed  by the formation of a ``frustrated'' network of cosmic strings --  $N$ and the inflationary energy scale are largely uncorrelated (see e.g. \cite{Burgess:2005sb}). This is admittedly an extreme example,  and we may explore constraints arising from the connection between $N$ and the inflationary scale  in future work.}  Roughly speaking, however, imposing $N>30$ is equivalent to setting the inflationary scale at or above the TeV scale.   This ensures that inflationary energy scales are not probed by precision electroweak measurements, and allows  baryogenesis to be driven by electroweak physics.  Given that the GUT scale is often regarded as the ``natural'' inflationary scale, assuming that inflation happens above the TeV scale is a very mild theoretical prior. However, demanding $N>30$ leads to very tight bounds on $\xi$. Translating our results into bounds on $\frac{dn_s}{d\ln{k}}$ yields a tighter constraint than that of Seljak {\em et al.} \cite{Seljak:2006bg} when we add the $N>30$ prior.  When we drop the constraint on $N$the central value of $\xi$ is unexpectedly large, and $\xi = 0$ is apparently excluded at the $2\sigma$ level by the WMAPII+SDSS dataset. When we compute $N$ for these chains we find a distribution peaked at 10-15 e-folds, consistent with   \cite{Easther:2006tv}, but in conflict with successful cosmological inflation.     We then turn to the pivot dependence of the bounds on the inflationary and spectral parameters, and show that in the absence of a prior on $N$, choosing a different pivot scale (comparing constraints at $0.002 \mbox{ Mpc}^{-1}$ to those at $0.02 \mbox{ Mpc}^{-1}$, in this case) somewhat tightens the bounds on $\xi$ and $\frac{dn_s}{d\ln{k}}$, while shifting $n_s$ significantly to the red. In the case with a ``sufficient e-folds" prior, the pivot dependence of the constraints is not strong. We stress that these results must be interpreted with caution, and make no claims to have found strong evidence for a non-zero value of $\xi$, as we have not yet made use of the full cosmological dataset. However, these results help us understand the significance of the weak preference for a negative running seen in the WMAPII dataset, and give us cause for optimism that future data will allow us to tightly constrain the inflationary parameter space.

This paper is arranged as follows. In Section 2 we briefly review the slow roll hierarchy, and its use in Monte Carlo Markov Chain analyses. Section 3 contains the constraints on the inflationary parameter space derived from our chains. In Section 4 we give our interpretation and analysis, including a discussion of the pivot-dependence of our constraints on the inflationary parameter space.  

\section{Imposing A Single-Field Slow-Roll Inflationary Prior}

The dynamics of single field inflation can be written in the Hamilton-Jacobi form, where overdots correspond to time derivatives and primes denote derivatives with respect to $\phi$  \cite{kinney:2002}.   The scale dependence of the  HSR parameters $^{\ell}\lambda_H$ is given by 
\begin{equation}
\epsilon(\phi) \equiv \frac{m^2_{\rm Pl}}{4\pi}
\left[\frac{H'(\phi)}{H(\phi)}\right]^2; \label{eq:eps} 
\end{equation}
\begin{equation}
^{\ell}\lambda_H \equiv \left(\frac{m^2_{\rm Pl}}{4\pi}\right)^\ell
  \frac{(H')^{\ell-1}}{H^\ell} \frac{d^{(\ell+1)} H}{d\phi^{(\ell+1)}}
   ;\ \ell \geq 1.  \label{eq:hier}
\end{equation}
The usual slow roll parameters are $\eta = {}^{1}\lambda_H$ and $\xi =   {}^{2}\lambda_H$.  If all terms with $\ell >M$ are zero at some fiducial point, these differential equations ensure they vanish at all other times. Liddle showed that the hierarchy can be solved exactly when truncated at order $M$ \cite{liddle:2003}, so  
\begin{equation}
\frac{H(\phi)}{H_0} = 1+ B_1 \left(\frac{\phi}{\mpl}\right) + \cdots + B_{M+1}
\left(\frac{\phi}{\mpl}\right)^{M+1} \, . \label{eq:h}
\end{equation}
The $B_i$ are specified by the initial values of the HSR parameters,
%%%%%%%%%%%%%%%%%%%%%%%%%%%%%%%%%%%%%%%%%%%%%%%%%%%
\begin{equation}
B_1 = \sqrt{4\pi\epsilon_0} \, \quad
B_{\ell+1} = \frac{(4\pi)^\ell }{(\ell+1)! 
\ B_1^{\ell-1}}  {}^{\ell}\lambda_{H,0}  \label{eq:coeffs}
\end{equation}
where the subscript $0$ refers to their values at the moment the fiducial mode $k_0$ leaves the horizon, and $\phi=\phi_0=0$. We choose $k_0 = 0.002$ Mpc$^{-1}$. At this point, we emphasize that this fiducial scale is not a pivot point - it ties the potential directly to a physical scale in the universe, and the closed nature of the truncation of the flow hierarchy then allows us to compute the primordial spectrum at any other physical scale. The potential is 
\begin{equation}
V(\phi) = \frac{3\mpl^2}{8\pi}    H^2\left(\phi\right)\left[1-\frac{1}{3}\epsilon\left(\phi\right)\right]  . \label{eq:v} 
\end{equation}
while  $N$ is given by 
\begin{equation}
\frac{dN}{d\phi} = \frac{4\pi}{\mpl^2} \frac{H}{H'} \, ,  \label{dndphi}
\end{equation}
and $\phi$ and $k$ are related by
\begin{equation}
\frac{d\phi}{d\ln k} = -\frac{\mpl}{2\sqrt{\pi}}
\frac{\sqrt{\epsilon}}{1-\epsilon}. \label{eq:phieq}
\end{equation}
%%%%%%%%%%%%%%%%%%%%%%%%%%%%%%%%%%%%%%%%%%%%%%%%%%%
Finally, we express the scalar and tensor primordial power spectra in terms of these parameters. A first order expansion around an exact solution for the case of power law inflation \cite{lidsey/etal:1995} gives
%%%%%%%%%%%%%%%%%%%%%%%%%%%%%%%%%%%%%%%%%%%%%%%%%%%
\begin{eqnarray}
P_{\cal{R}} &=&\left.  \frac{\left[1-(2C+1)\epsilon +
C\eta\right]^2}{\pi\epsilon}\left(\frac{H}{\mpl}\right)^2 \right|_{k=a H},
\label{eq:Pscalar} \\
P_h &=&\left. \left[1-(C+1)\epsilon\right]^2
\frac{16}{\pi}\left(\frac{H}{\mpl}\right)^2 \right|_{k=a H},
\label{eq:Ptensor}
\end{eqnarray}
%%%%%%%%%%%%%%%%%%%%%%%%%%%%%%%%%%%%%%%%%%%%%%%%%%%
where $\eta =\ ^{1}\lambda_H$, $C= -2 + \ln{2} + \gamma \approx -0.729637$ and $\gamma$ is the Euler-Mascheroni constant.   One minor inconsistency in our approach is that (\ref{eq:Pscalar}) and (\ref{eq:Ptensor}) are calculated using the slow roll approximation, and truncated at second order.  In practice,  we do not believe this is a significant drawback, as the datasets we are working with provide meaningful constraints on (at most) the lowest three slow-roll parameters. In the context of future high-precision data, this problem could be surmounted by either using a higher order expansion for the spectra, or even integrating the mode equations directly.

We can write the usual observables in terms of the slow roll parameters, 
\begin{eqnarray}
n_s &=& 1+ 2\eta - 4\epsilon - 2(1+{\cal C}) \epsilon^2 - \frac{1}{2}(3-5{\cal C}) \epsilon \eta + \frac{1}{2}(3-{\cal C})\xi   \label{eq:hsrconversionsstart} \\
r &=& 16 \epsilon \left[1+2  C(\epsilon - \eta)\right]\\
 \frac{dn_s}{d\ln{k}} &=& -\frac{1}{1-\epsilon} \left\{2 \xi + 8 \epsilon^2 - 10 \epsilon \eta +  \frac{7{\cal C}-9}{2} \epsilon \xi    + \frac{3-{\cal C}}{2} \xi \eta \right\}  \label{eq:nsrun} \\
n_t &=& -2\epsilon - (3+{\cal C}) \epsilon^2 + (1+{\cal C}) \epsilon \eta 
\end{eqnarray}
where ${\cal C}=4(\ln 2 + \gamma) - 5$, and we have introduced the customary notation $\alpha  = dn_s/ d\ln{k}$; $r$ is the tensor/scalar ratio\footnote{Beware of the distinction between ${\cal C}$ and $C$ -- these quantities are sometimes confused in the literature.} and $n_t$ is the tensor spectral index. We retain all terms in $\frac{dn_s}{d\ln{k}}$ up to quadratic order in the slow roll parameters, anticipating that $\xi$ may be as large as $\epsilon$ or $\eta$. Note that these are {\em derived} quantities and are never directly used in the calculation of power spectra in this work. 

\FIGURE[!htb]{
\includegraphics[width=2.5in]{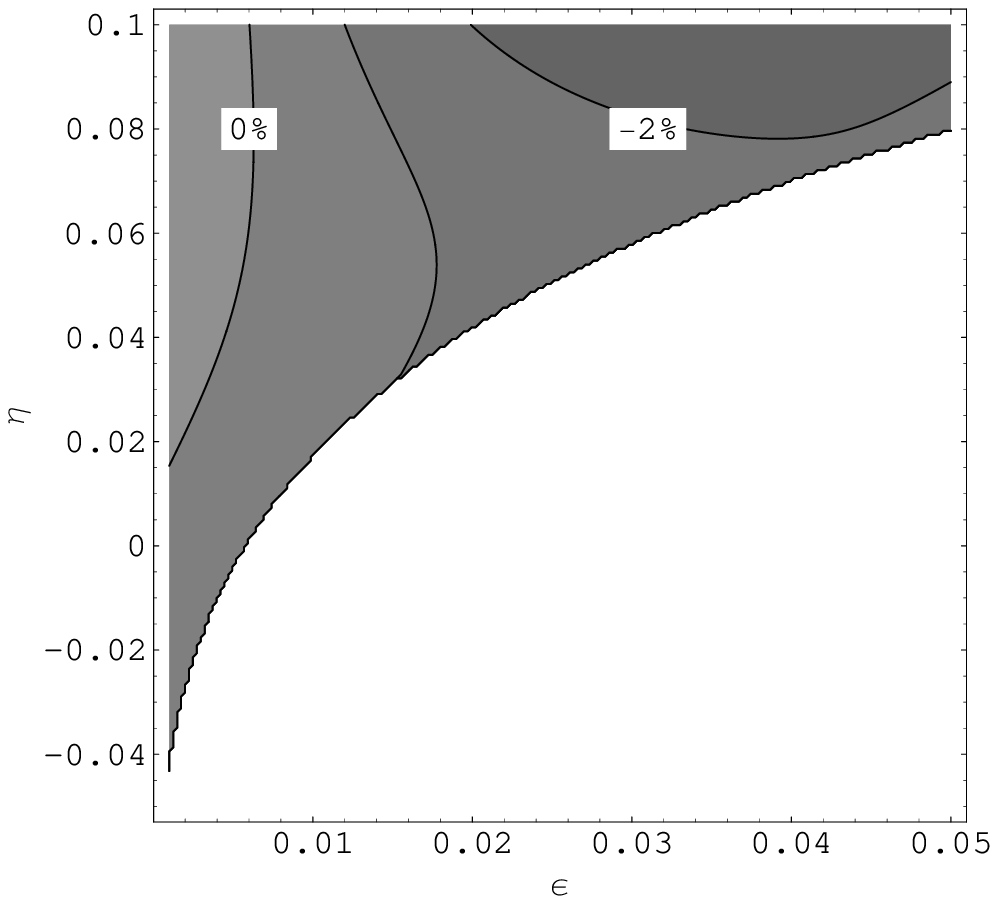} 
\includegraphics[width=2.5in]{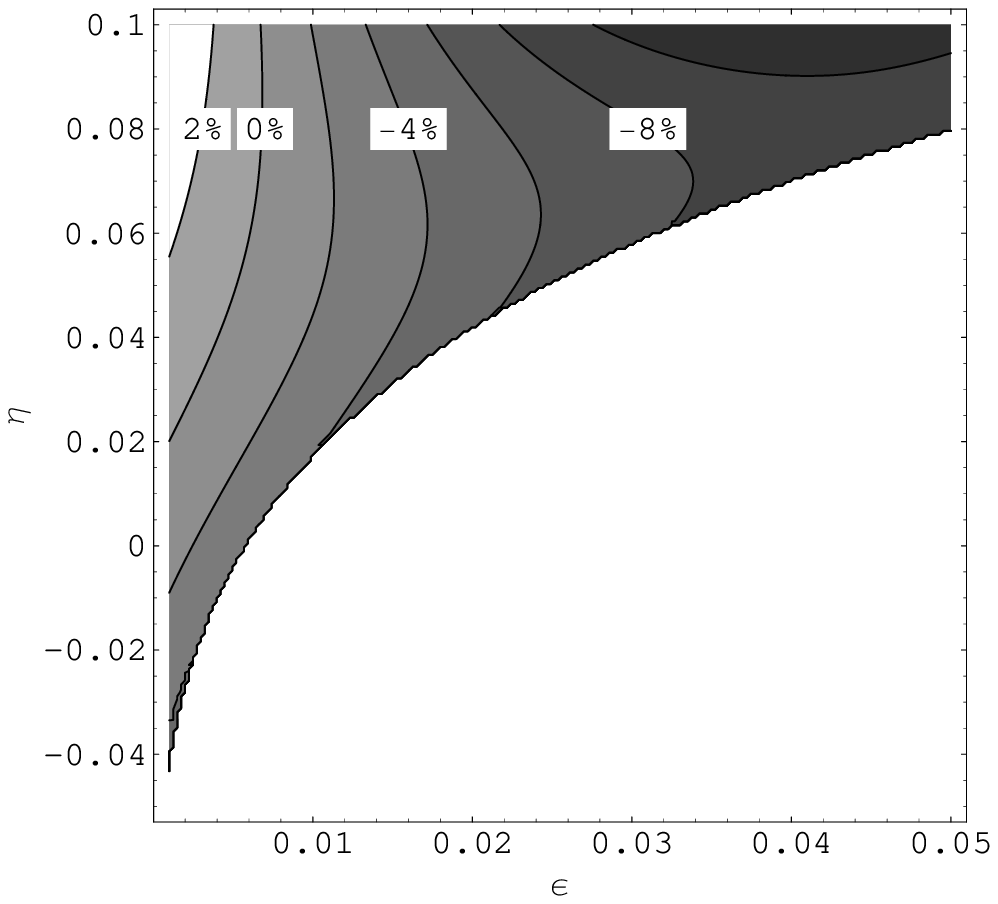} 
\caption{ The difference between the scalar power spectrum computed using $n_s(\epsilon,\eta,\xi)$ and the full HSR parameters at 0.1 Mpc$^{-1}$ (left) and 1 Mpc$^{-1}$ (right), for $\xi = 0.01$. A  positive difference corresponds to the HSR result being larger than the $n_s(\epsilon,\eta,\xi)$  value.  The ``missing'' region corresponds to parameter choices which would lead to less than 15 additional e-folds of inflation. Note that  $\xi = 0.01$ is significantly smaller than the central value implied by the data, and the discrepancy grows with $\xi$.} \label{fig:accuracy}
}
 
It is also worth emphasizing that the inflationary single-field consistency condition $n_t=-r/8$ only strictly holds at the pivot point when the spectra are specified in terms of their respective indices and running.  This is easily seen by writing out the scale dependence of the standard power-law variables:
%%%%%%%%%%%%%%%%%%%%%%%%%%%%%%%%%%%%%%%%%%%%%%%%%%%%%%%%%%%%%%%%%%%
\begin{eqnarray}
  P_{\cal R}(k)&=& P_{\cal R}(k_0)
  \left(\frac{k}{k_0}\right)^{n_s(k_0)-1+\frac{1}{2}(dn_s/d\ln   k)\ln(k/k_0)},  \label{eq:pkempirical} \\ 
  P_h(k)&=& P_h(k_0)
  \left(\frac{k}{k_0}\right)^{n_t(k_0)+\frac{1}{2}(dn_t/d\ln
  k)\ln(k/k_0)}. \label{eq:pk_scale_dep}
\end{eqnarray}
%%%%%%%%%%%%%%%%%%%%%%%%%%%%%%%%%%%%%%%%%%%%%%%%%%%%%%%%%%%%%%%%%%%
Hence, while it is common to assume that $dn_t/d\ln k =0$ and $n_t = -r(k_0)/8$ is constant, the scale-dependence of the tensor/scalar ratio $r(k) = P_h(k)/P_{\cal R}(k)$ is given by
%%%%%%%%%%%%%%%%%%%%%%%%%%%%%%%%%%%%%%%%%%%%%%%%%%%%%%%%%%%%%%%%%%%
\begin{equation}
  r(k)= r(k_0)
  \left(\frac{k}{k_0}\right)^{n_s(k_0)-1+\frac{1}{2}(dn_s/d\ln
  k)\ln(k/k_0)+r(k_0)/8} \label{eq:r_k} 
\end{equation}
%%%%%%%%%%%%%%%%%%%%%%%%%%%%%%%%%%%%%%%%%%%%%%%%%%%%%%%%%%%%%%%%%%%
and it is apparent that the consistency condition does not hold away from the pivot. This is an unsatisfactory state of affairs if one wishes to impose a single-field slow-roll prior, and slow roll reconstruction does not suffer from this shortcoming.

An alternative to slow roll reconstruction is  given by the work of Leach {\em et al.\/} \cite{Leach:2002ar,Leach:2002dw,Leach:2003us,Finelli:2006fi},  who specify the spectrum in  terms of the slow roll parameters, but then compute its scale dependence via equations \ref{eq:hsrconversionsstart} to \ref{eq:nsrun}.   Conversely, we dispense with an explicit pivot point and compute the spectra directly from equations \ref{eq:Pscalar} and \ref{eq:Ptensor}, while using the flow equations to obtain the appropriate scale dependent values of the slow roll parameters.  When only $\epsilon$ and $\eta$ are included in the analysis, these two approaches (i.e. ours vs. Leach  {\em et al.\/}'s) overlap to with a few percent over the full observable range.  However, when $\xi$ is included the discrepancy can be much larger. For instance with a pivot at $k=0.002\mbox{ Mpc}^{-1}$, the spectra computed using the formalism of \cite{Leach:2002ar,Leach:2002dw,Leach:2003us} differ from that obtained using the flow equations by as much as 10\% at  very short scales, as illustrated in Figure \ref{fig:accuracy}. This discrepancy could be corrected by extending the approach of \cite{Leach:2002ar,Leach:2002dw,Leach:2003us}  to higher order in slow roll. 
 
We perform standard MCMC fits to cosmological data, using modified versions of CAMB\footnote{{\tt http://camb.info/}} \cite{Lewis:1999bs} to provide the CMB spectra and {\sc CosmoMC}\footnote{{\tt http://cosmologist.info/cosmomc/}} \cite{Lewis:2002ah} to handle the Markov chains. All our chains  include uniform priors on the slow roll parameters $\{\epsilon_0,\eta_0,\xi_0\}$, where $\epsilon_0$, $\eta_0$, and $\xi_0$ are the values at $k_0 = 0.002$ Mpc$^{-1}$. The primordial curvature power spectrum is normalized at $k_0$ by setting
%%%%%%%%%%%%%%%%%%%%%%%%%%%%%%%%%%%%%%%%%%%%%%%%%%%
\begin{equation}
A_s = \frac{\left[1-(2C+1)\epsilon_0 +
C\eta_0\right]^2}{\pi\epsilon_0}\left(\frac{H_0}{\mpl}\right)^2, \label{eq:norm}
\end{equation}
%%%%%%%%%%%%%%%%%%%%%%%%%%%%%%%%%%%%%%%%%%%%%%%%%%%
and the uniform prior is on $\ln[10^{10} A_s]$. Hereafter, we drop the subscripts $0$ for brevity. The  late time parameters are $\Omega_b h^2$, $\Omega_{CDM} h^2$, $h$ (the uniform prior is on $\theta_A$) and $\tau$, where these variables have their usual meanings.  Since we impose an inflationary prior we take $\Omega_{\mbox{\tiny Total}} = 1$ so the contribution from dark energy is $\Omega_\Lambda = 1- \Omega_b - \Omega_{CDM}$, and we set $w_\Lambda=-1$. Instantaneous reionization is assumed. To constrain this parameter set, WMAPII and SDSS data are used. We use the $v2p1$ version of the WMAP 3 year likelihood code\footnote{{\tt http://lambda.gsfc.nasa.gov/product/map/dr2/}} with the same options that were employed in \cite{Peiris:2006ug}, and the SDSS \cite{Tegmark:2003uf} likelihood routine provided in the May 2006 version of {\sc CosmoMC}. The latter marginalizes over the normalization bias of the SDSS galaxy power spectrum with respect to the underlying matter power spectrum. We do not impose an age prior, nor do we employ the HST prior for the Hubble constant. In each case we ran 8 individual chains.

\TABLE[!t]{
 \begin{tabular}{||c|c|c||}
\hline
  Chain \# & Dataset Combination & E-folds Constraint   \\
\hline 
1. &  WMAPII +SDSS  & --  \\
\hline
2. & WMAPII  & $N>30$    \\
\hline  
3. & WMAPII +SDSS &  $N>30$\\
\hline
\end{tabular}
 \caption{ MCMC Parameter fits  described in this paper.   
}
\label{table:summary}
}

\TABLE[!t]{
\begin{tabular}{||c|c|c|c||}
\hline
Parameter           & WMAPII+SDSS   	   &  WMAPII             &  WMAPII+SDSS \\
		    & (no $N$ restriction) & ($N>30$) 	         & ($N>30$) \\
\hline 
$\epsilon$          & $<0.013$ (95\% CL)    & $<0.033$  (95\% CL)  &  $<0.015$ (95\% CL)\\
\hline 
$\eta$      	    & $0.029 \pm 0.021$    & $0.024^{+0.029}_{-0.028}$  &  $0.0 \pm 0.014$ \\
\hline
$\xi$               & $0.029 \pm 0.013$    & $0.0 \pm 0.003$    &  $0.0 \pm 0.002$ \\
\hline
$\ln[10^{10} A_s]$  & $3.17 \pm 0.06$      & $3.03 \pm 0.10$    &  $3.17 \pm 0.06$ \\
\hline
\end{tabular}
\caption{Constraints on HSR parameters defined at $k=0.002$ Mpc$^{-1}$. Constraints are at the 68\% confidence level unless otherwise noted.}\label{table:hsr_constraints}}

%WMAP+SDSS no xi DIFFERENT SDSS LIKELIHOOD NORMALIZATION  5641.03320312500000
%WMAP+SDSS no prior 5635.55175781250000
%WMAP+SDSS N>30	   5637.22119140625000
%WMAP      N>30     5624.23486328125000

\TABLE[!t]{
\begin{tabular}{||c|c|c|c||}
\hline
Parameter           & WMAPII+SDSS   	   &  WMAPII             &  WMAPII+SDSS \\
		    & (no $N$ restriction) & ($N>30$) 	         & ($N>30$) \\
\hline 
$n_s$               & $1.08 \pm 0.05$    & $0.99 \pm 0.02$    &  $0.98 \pm 0.02$ \\
\hline 
$d n_s/d\ln k$      & $-0.057^{+0.027}_{-0.026}$   & $0.004 \pm 0.006$   &  $0.0^{+0.004}_{-0.003}$ \\
\hline
$r$                 & $<0.22$ (95\% CL)   & $<0.56$ (95\% CL)  &  $<0.24$ (95\% CL) \\
\hline
$n_t$               & $>-0.026$ (95\% CL)  & $>-0.067$ (95\% CL)  &  $>-0.030$ (95\% CL) \\
\hline
\end{tabular}
\caption{Constraints on derived primordial parameters defined at $k=0.002$ Mpc$^{-1}$. Constraints are at the 68\% confidence level unless otherwise noted.}\label{table:derived_constraints}}

\section{Results}

We have run three different sets of chains, as summarized in Table~\ref{table:summary}. We estimate the parameters for WMAPII alone, and for WMAPII+SDSS, with no constraint on $N$ and with the requirement that $N>30$.  In each case we tested convergence via the usual Gelman \& Rubin \cite{gelman/rubin:1992} criterion.  The chains that drew only on the WMAPII dataset took a very long time to converge. Without the $N>30$ constraint we were not close to convergence even after taking over 500,000 accepted steps. For this reason we do not report constraints for the WMAPII dataset on its own, without a prior on $N$.   When SDSS data was included, convergence was comparatively prompt, reflecting the greater leverage provided by the combination of CMB and LSS data.  

The prior (or lack thereof) on $N$ is implemented as follows. Each draw of $\{\epsilon, \eta, \xi\}$ in the chains maps uniquely onto a potential. For this potential, we can evolve it forward in time from when the fiducial scale $k_0$ leaves the horizon and compute when $\epsilon=1$, corresponding to the end of inflation. In the case with no prior on $N$, we only check that inflation lasts for at least the few e-folds encompassing the CMB and LSS data that we actually use -- if inflation ends within this observable window, that model is discarded. For the case where $N>30$, we require that the potential so specified provides at least an additional 30 e-folds of inflation after the fiducial scale leaves the horizon. Otherwise it is rejected from the chain.

\FIGURE[!tb]{ 
\includegraphics[width=4.3in]{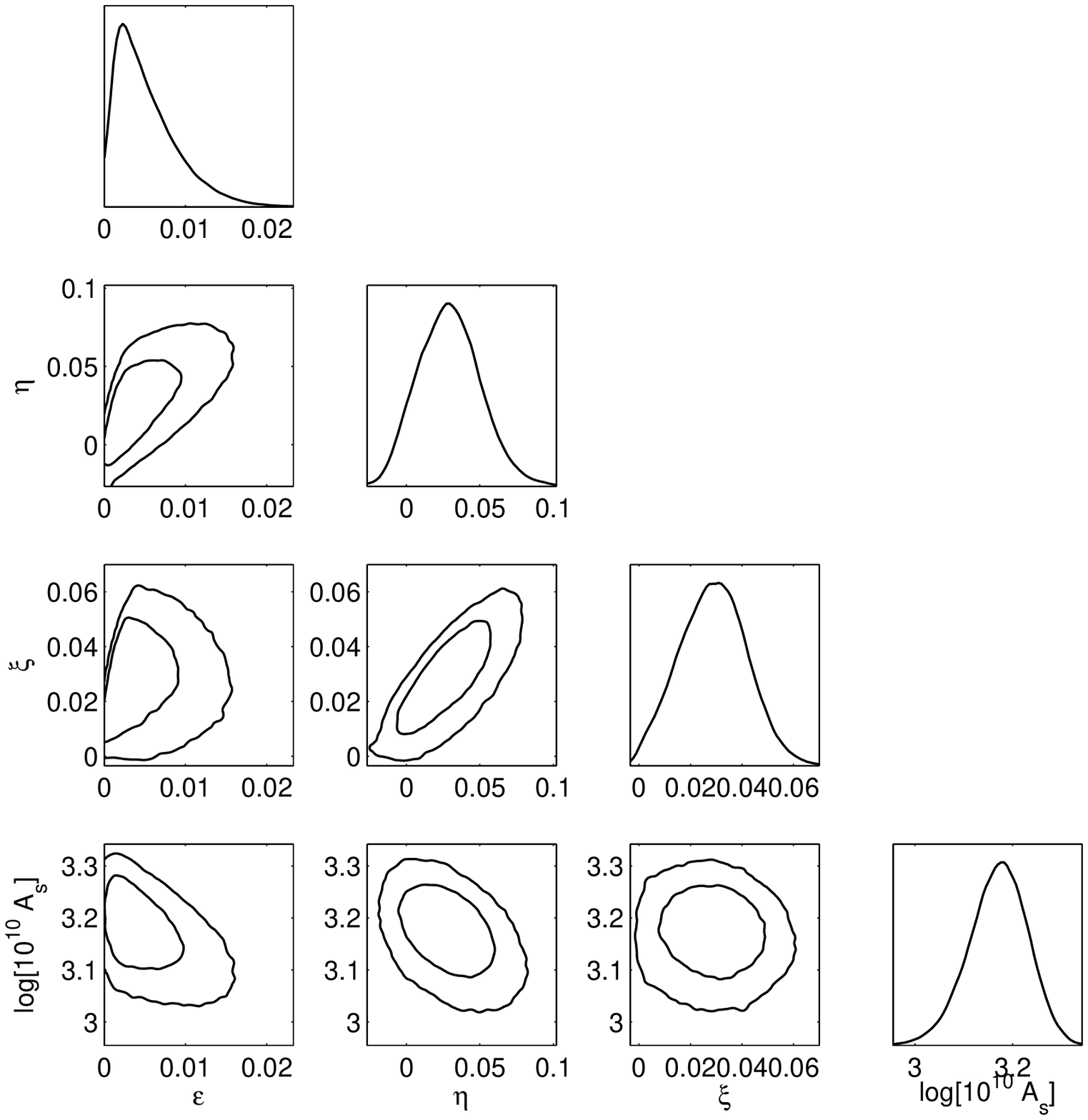}
\caption{Primordial parameter constraints for WMAPII+SDSS (no $N$ constraint). The top plot in each column shows the probability distribution function for each of the parameters, while the other plots show their joint 68\% and 95\% confidence levels.} \label{fig:WMAPII-SDSS}
}

\FIGURE[!tb]{ 
\includegraphics[width=4.3in]{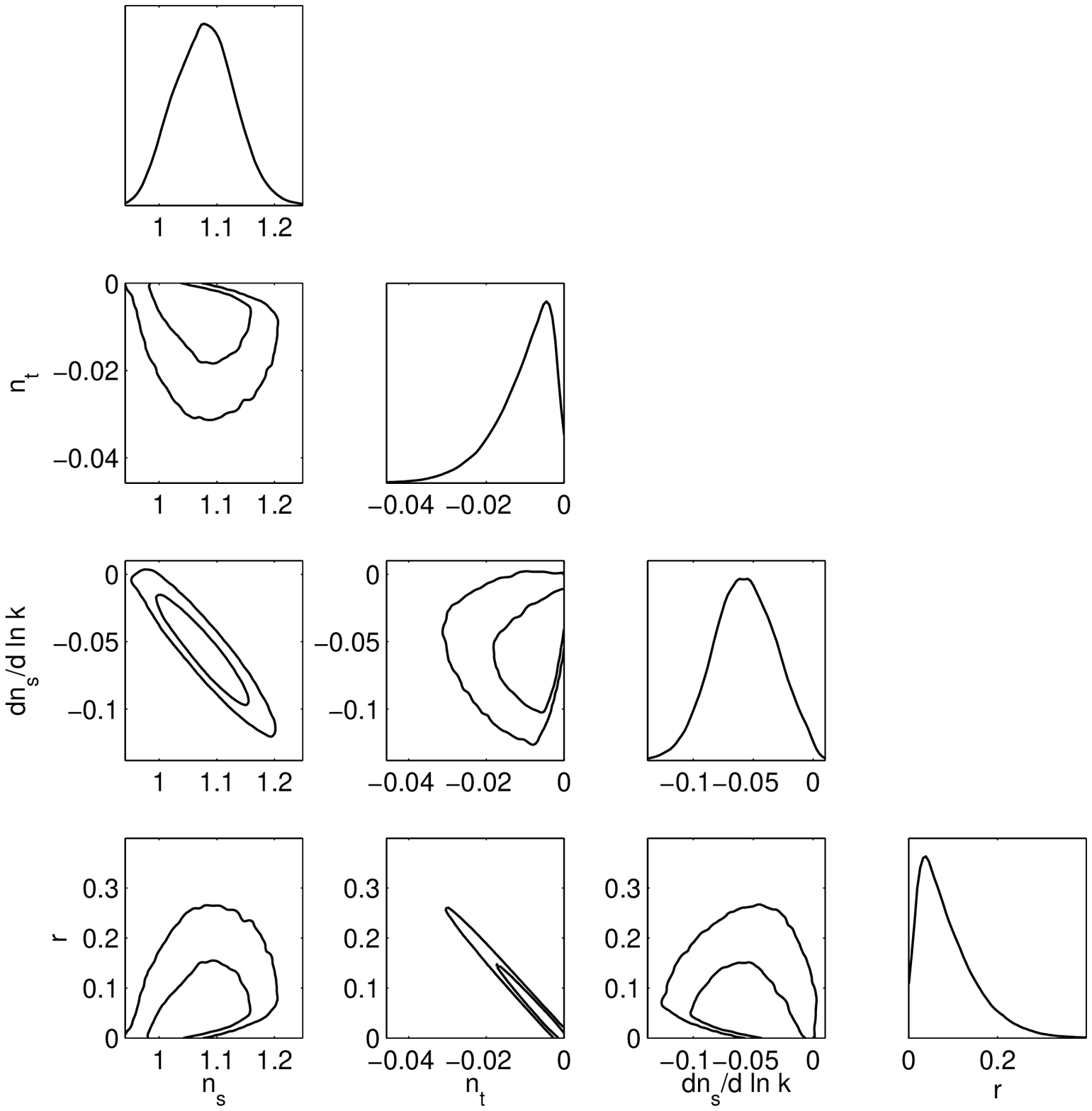}
\caption{Primordial parameter constraints for WMAPII+SDSS (no $N$ constraint) in conventional power law variables. The top plot in each column shows the probability distribution function for each of the parameters, while the other plots show their joint 68\% and 95\% confidence levels. The power law variables are derived parameters which are not used in the MCMC chains themselves.} \label{fig:WMAPII-SDSS-std}
}

\FIGURE[!tb]{ 
\includegraphics[width=4.3in]{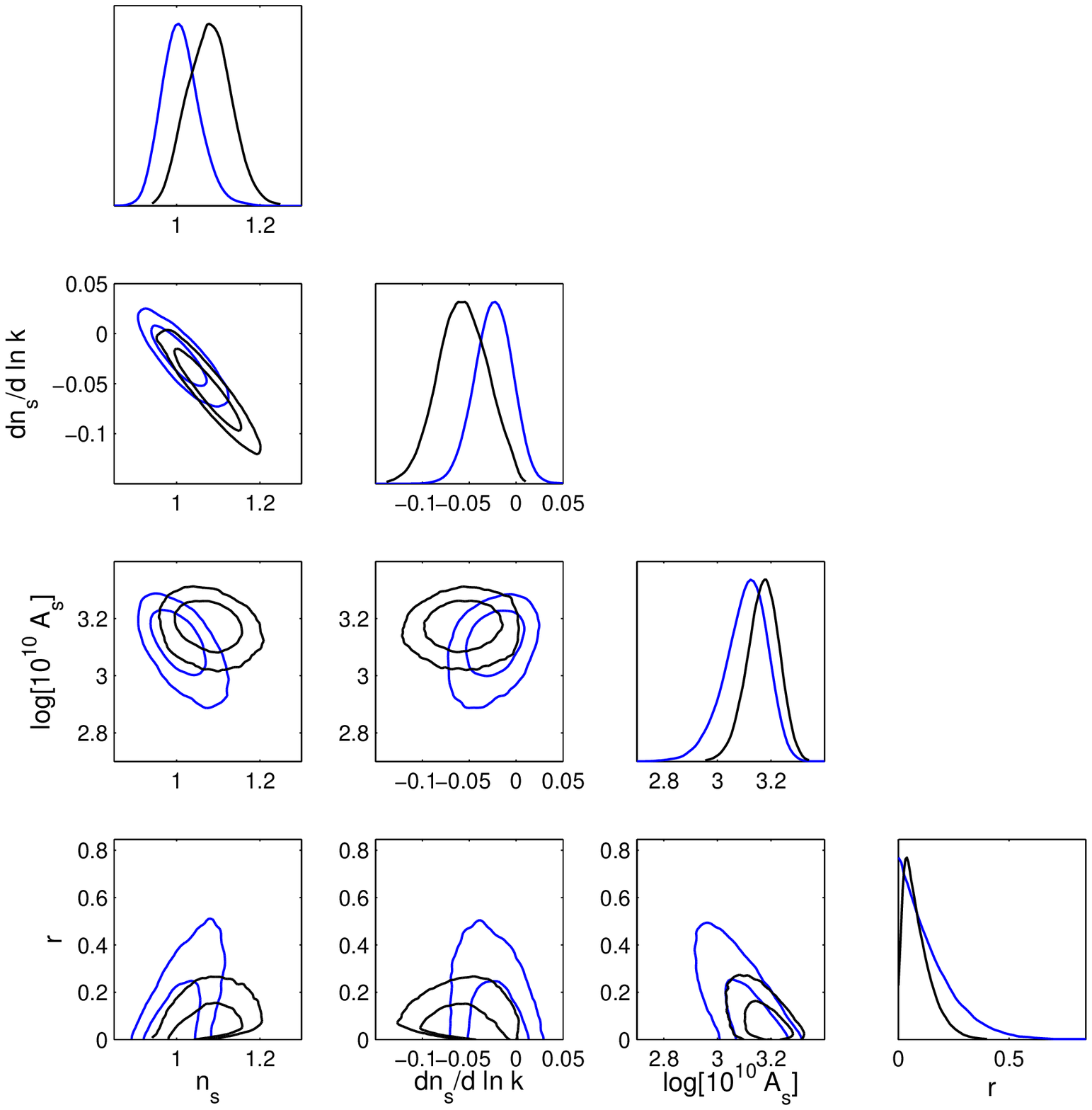}
\caption{Primordial parameter constraints for WMAPII+SDSS (no $N$ constraint) in conventional power law variables (black), compared with the most directly comparable results from \cite{Spergel:2006hy} using chains obtained from {\tt http://lambda.gsfc.nasa.gov} (blue). The Spergel et al. chains contain an additional bias constraint on the SDSS data which our runs did not include.} \label{fig:WMAPII-SDSS-comp}
}

\FIGURE[!tb]{ 
\includegraphics[width=4.3in]{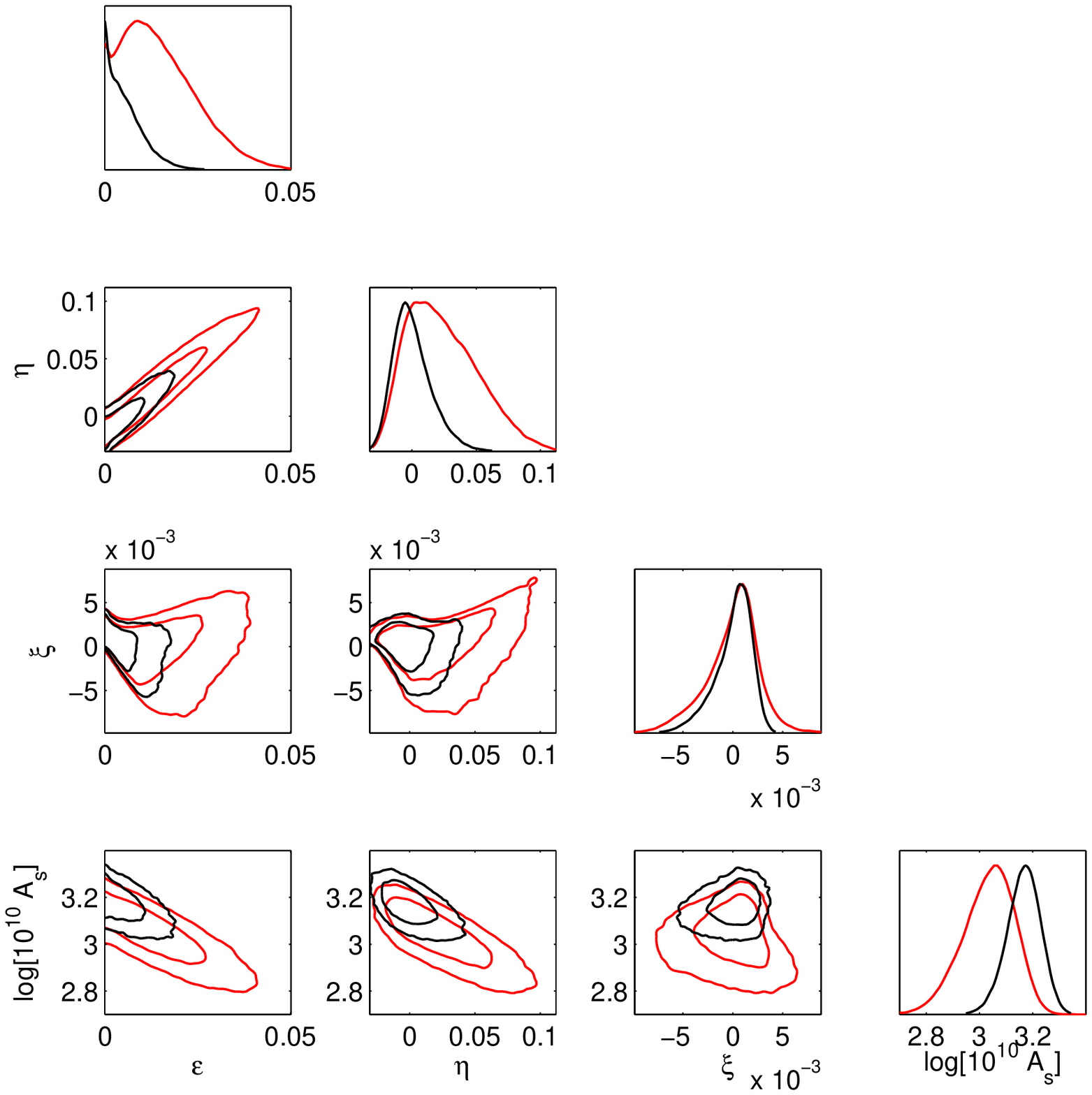}
\caption{Primordial parameter constraints for WMAPII+SDSS, (black) and WMAPII (red), imposing a prior $N>30$. The top plot in each column shows the probability distribution function for each of the parameters, while the other plots show their joint 68\% and 95\% confidence levels.} \label{fig:Nconstraint}
}

\FIGURE[!tb]{ 
\includegraphics[width=4.1in]{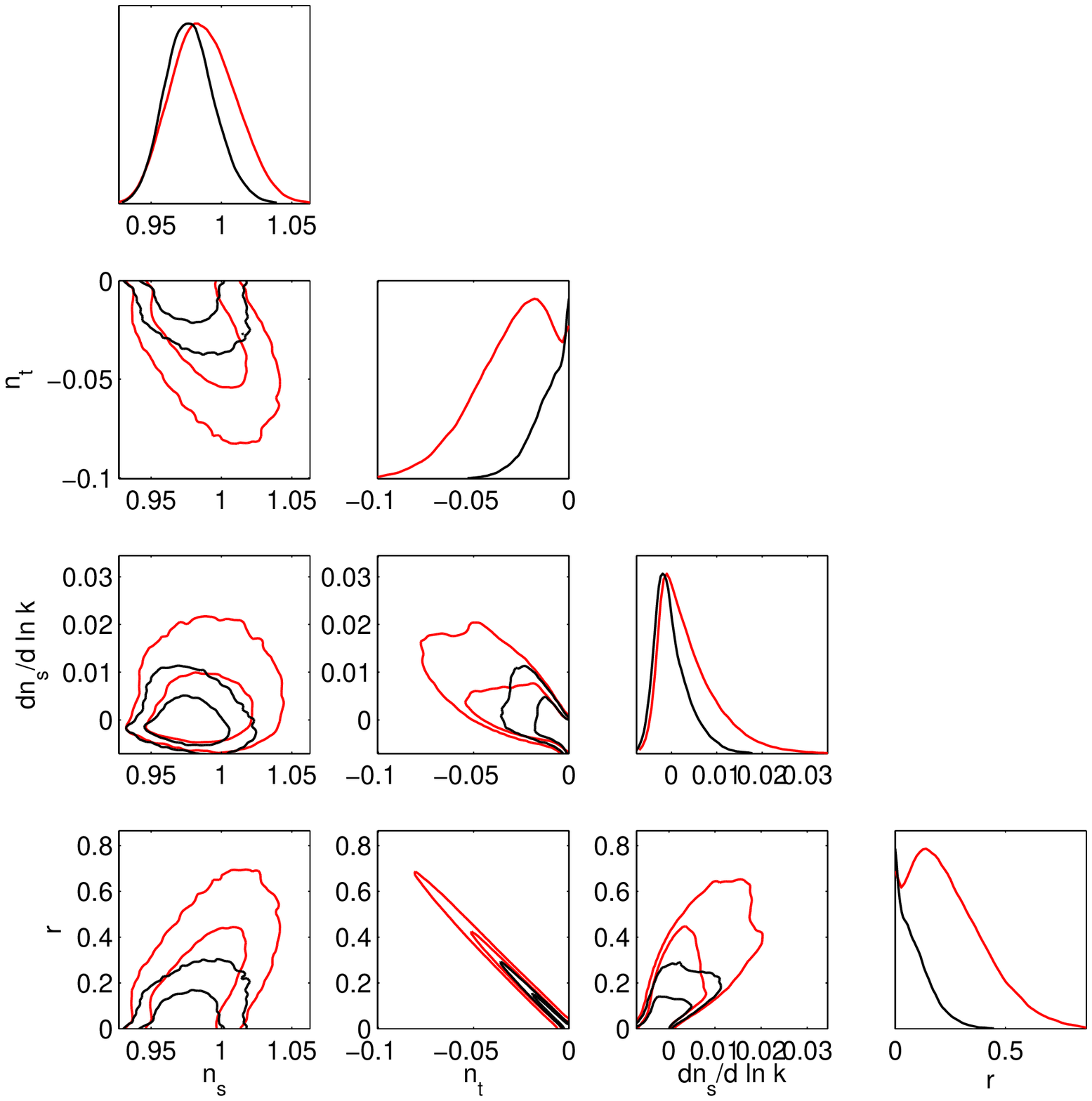}
\caption{Primordial parameter constraints for WMAPII+SDSS, (black) and WMAPII (red), imposing a prior $N>30$, in conventional variables. The top plot in each column shows the probability distribution function for each of the parameters, while the other plots show their joint 68\% and 95\% confidence levels. The power law variables are derived parameters which are not used in the MCMC chains themselves.} \label{fig:Nconstraint-std}
}

The parameter bounds derived from each set of chains are given in Tables~\ref{table:hsr_constraints} and \ref{table:derived_constraints}, while Figures~\ref{fig:WMAPII-SDSS} through \ref{fig:Nbound} show the results derived from the chains we ran.    In particular, Figure~\ref{fig:WMAPII-SDSS} shows the limits on the slow roll parameters derived from Set 1, where we are fitting to the combined power spectrum derived from WMAPII and SDSS, with no constraint on the number of e-folds.  In Figure~\ref{fig:WMAPII-SDSS-std} we show the same results, after they have been mapped into the ``standard'' slow roll variables.   These plots are analogous to those shown in  \cite{Peiris:2006ug}, except for the presence of the $\xi$ parameter.  Comparing these plots with those from our previous paper, we see that including $\xi$ greatly increases the allowed range of values of the running, compared to the $\{\epsilon,\eta\}$ chains. Indeed, a naive interpretation of the data suggests that there is the evidence  for $\xi>0$ beyond the $2\sigma$ level -- but there are a number of issues surrounding this result, and we stress that we make no claim to have found a non-trivial value of $\xi$.    In Figure~\ref{fig:WMAPII-SDSS-comp} we compare these results to chains run by Spergel {\em et al.\/}   \cite{Spergel:2006hy} for the same dataset. The two sets of chains are not exactly equivalent as the Spergel results have an additional constraint on the bias for the SDSS, which we do not include in the chains described here. We again see that a naive inspection of the allowed portion of the parameter space shows that working with an inflationary prior imposed via the use of the slow roll parameters in the chains leads to a stronger preference for a negative running than was found by Spergel {\em et al.\/}    A further difference is that upper bound on the tensor contribution found using a slow roll prior is considerably lower than that obtained from the $\{n_s,r,\frac{dn_s}{d\ln{k}}\}$ parametrization of the spectrum. This phenomenon was also observed in  \cite{Peiris:2006ug} in the absence of the running, and appears to be related to the fact that the degeneracy between $\epsilon$ and $\eta$  in terms of their contribution to $n_s$ is broken by the way these quantities contribute to the running index and $r$.    

 A further, more general, message to be drawn from  the two sets of chains shown in Figure~\ref{fig:WMAPII-SDSS-comp} is that the strength of the evidence for a running spectrum is dependent on the way we parameterize the running.  One might assume that imposing a strong inflationary prior by directly including the slow roll parameters in the chains would decrease the available region of parameter space, relative to that found from using the conventional $\{n_s,r,\frac{dn_s}{d\ln{k}}\}$ formalism.  However, we see that the two parameter volumes are slightly offset, and neither one is a strict subset of the other.   This reminds us that parameterizing the power spectrum in terms of the index and running is simply an empirical fit to what we assume is a weakly scale dependent function, but this characterization of the spectrum is not fundamental. A further caveat is that we do not impose a bias prior for the SDSS power spectrum in our chains, unlike those of \cite{Spergel:2006hy}. This constraint could have a significant impact on the estimate of the running, since it effectively fixes the normalization of the SDSS spectrum. We cannot distinguish at present between which of these effects -- the change in priors and parameterization, or the SDSS bias prior -- is most responsible for this difference from \cite{Spergel:2006hy}. This would require re-running our model with the SDSS bias prior; we leave this issue for possible future study.

As noted above, using the slow roll prior puts tighter constraints on the amplitude of the tensor spectrum than the  $\{n_s,r,\frac{dn_s}{d\ln{k}}\}$  chains.  If one looks carefully  at Figures \ref{fig:WMAPII-SDSS} and \ref{fig:WMAPII-SDSS-std}, one sees that $\epsilon$ and $r$ are peaked at a non-zero value.  This result is not significant, but it is worth pointing out that if the potential does have a complicated shape, it may be possible to measure $\epsilon$ without first detecting the primordial tensor contribution to the CMB.   Recall that $\epsilon$ determines how fast the inflaton field evolves, whereas the scalar index is (to lowest order) a linear combination of $\epsilon$ and $\eta$.  Finally, we remind the reader that in the conventional parameterization, the slope and amplitude of the primordial tensor spectrum are  determined by $r$.   Consequently, the  plots in Figure \ref{fig:WMAPII-SDSS-comp} do not contain $n_t$, whereas in Figure~\ref{fig:WMAPII-SDSS} we can plot a distribution in the $\{n_t,r\}$ plane, since second order terms in slow roll ensure that $n_t$ is not strictly proportional to $r$.

A  novel feature of slow roll reconstruction is that we obtain information about the likely number of e-folds of inflation, estimated in terms of $\epsilon,\eta$ and $\xi$ measured at some fiducial moment during inflation.  Conversely, we can invert this process and insert a prior on the number of e-folds into the parameter estimation algorithm. In practice, we   only work   with lower limits  on $N$. While   it is not possible for an arbitrary amount of inflation to occur after cosmological scales leave the horizon,   there is a lower bound on the number of e-folds if the universe is to successfully reheat.   In many models, the end of inflation   occurs when an instability develops in a direction in field-space that is orthogonal to the inflationary trajectory. In that case there is no ``advance warning'' that inflation is about to end, and the slow roll parameters may remain small right up until the transition to non-inflationary expansion. Here a naive computation of  $N$ using the values the slow roll parameters as the CMB scales leave the horizon yields a misleadingly large number. Conversely, a low value of $N$ (which is practice means anything less than 30), implies  that inflation will end too quickly unless some further effect comes into play. This could be a contribution from higher order terms in slow roll, or  a second period of inflation -- both of which are excluded by our prior.  Consequently, we  can usefully include a lower bound on $N$ in our fits, but we do not impose an upper bound.

In Sets 2 and 3 we impose the prior $N>30$ on both WMAPII, and the combination  WMAPII+SDSS,  and plot the results in Figure \ref{fig:Nconstraint}.   These chains put much tighter limits on $\xi$ than   Set 1, which has no  constraint on $N$.  Further, the addition of the SDSS data to WMAPII significantly tightens the parameter constraints. Comparing our results here with Figure 5 of \cite{Peiris:2006ug}, where we perform a fit to the first two slow roll parameters, we see that the permitted region of the $\{\epsilon,\eta\}$ plane does not change substantially when the $\xi$ parameter is added along with the $N>30$ constraint. Without the $N>30$ constraint adding $\xi$ significantly expands the allowed  range of $\{\epsilon,\eta\}$.

In Figure~\ref{fig:Nconstraint-std} the two sets of $N>30$ chains are converted into constraints on $n_s,r$ and $\frac{dn_s}{d\ln{k}}$. The constraints on $\frac{dn_s}{d\ln{k}}$ are an order of magnitude tighter than those we obtained in the absence of the $N>30$ prior.   The tightest published constraints on $\frac{dn_s}{d\ln{k}}$ are those given by \cite{Seljak:2006bg}, who include Lyman-$\alpha$ and SN1a data, along with the information from SDSS and WMAPII that we employ here.  This combination yields $\frac{dn_s}{d\ln{k}} = (-1.5 \pm 1.2) \times 10^{-2}$, where the error range spans the  68\% confidence interval.  We present our constraints on the slow roll parameters in Table~\ref{table:hsr_constraints}, while Table~\ref{table:derived_constraints} contains the corresponding limits on the spectral parameters.    With the $N>30$ constraint, our bound on $\frac{dn_s}{d\ln{k}}$ is $6\times10^{-3}$ -- half that of \cite{Seljak:2006bg}, and it shrinks again when we include the SDSS data, and both analyses return a running that is peaked very close to zero. Conversely, our constraint on $n_s$ is noticeably broader, and shifted toward the blue, relative to that of  \cite{Seljak:2006bg} -- we shall see later that this may be due to the fact that they report their constraints at a pivot point ($0.05$ Mpc$^{-1}$) which is at significantly smaller scales than ours.  We do not include the dark energy equation of state ($w$) in our analysis, whereas this is a free parameter in  \cite{Seljak:2006bg}. The differences between our results and those of \cite{Seljak:2006bg} reflects different underlying assumptions, rather than any intrinsic ``conflict'' between the two analyses.  The authors of  \cite{Seljak:2006bg} set out to work with as large a range of data sources as possible, whereas our approach is to work with only one or two probes of the fundamental power spectrum, and to impose a strong theoretical prior -- namely slow roll inflation -- on the mechanism responsible for the generation of perturbations.

\FIGURE[!tb]{ 
\includegraphics[width=4in]{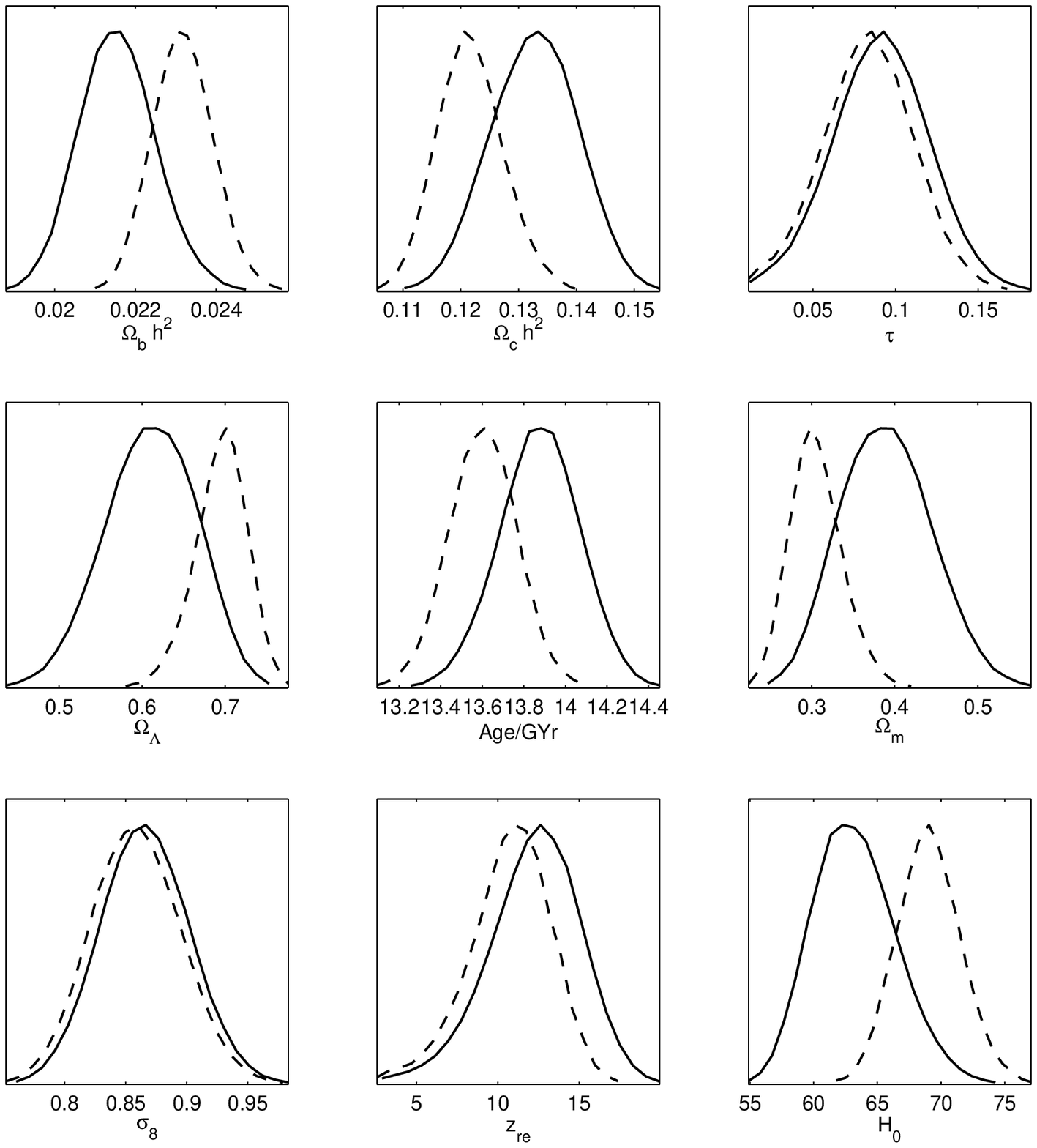}
\caption{Marginalized 1D posterior distributions for the ``late time variables'' for WMAPII+SDSS, imposing a prior $N>30$ (dashed) and with no constraint on $N$ (solid). $\tau$ and $\sigma_8$ are remarkably stable to the addition of the prior; most other parameters experience shifts in the mean at the level of a $\sigma$ or more. The greatest effect is on $\Omega_m$ (see the discussion in text for details).} \label{fig:lateparams}
}

\FIGURE[!tb]{ 
\includegraphics[width=4in]{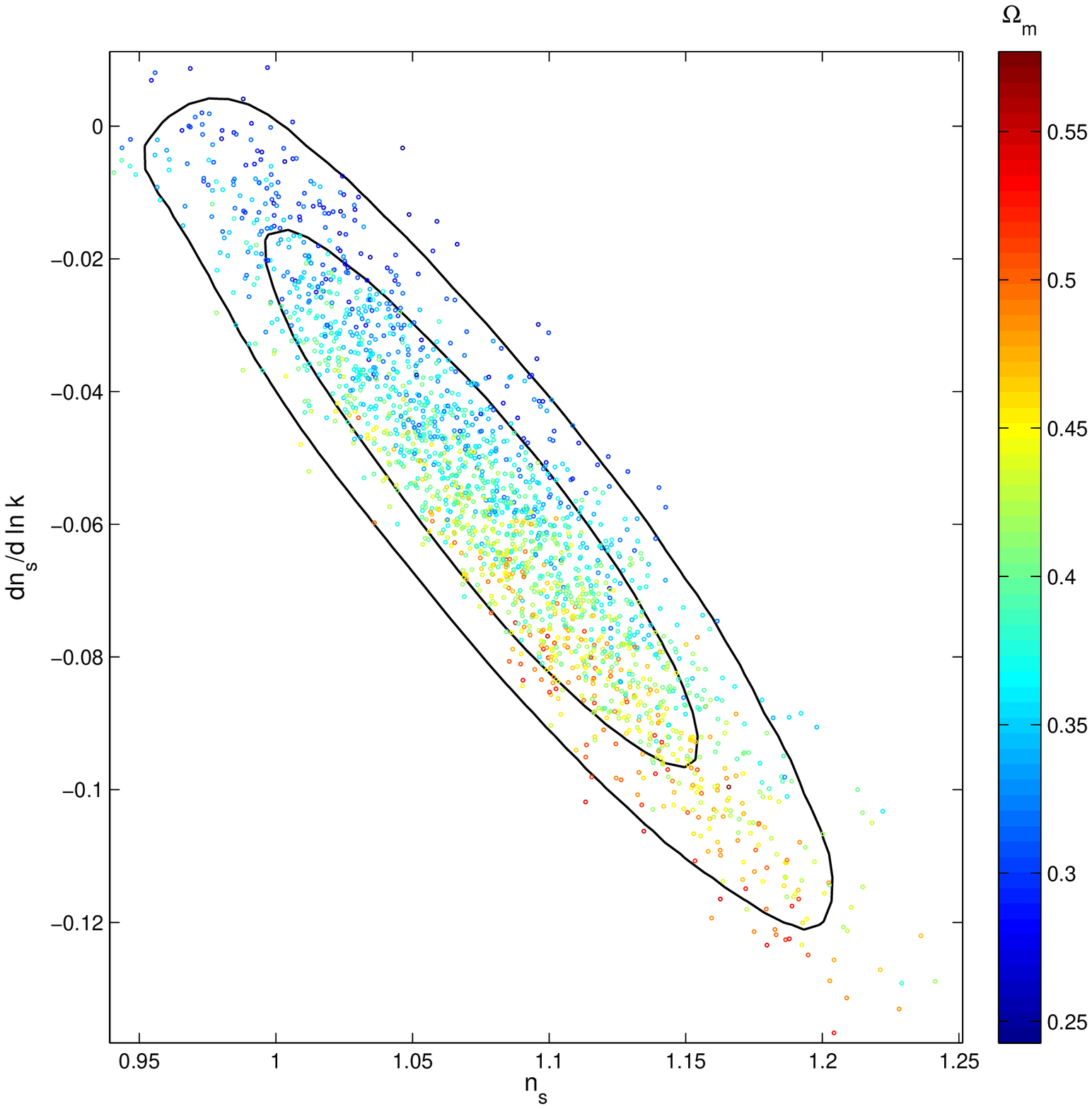}
\caption{The joint 68\% and 95\% confidence levels in the $n_s$ vs $d n_s/ d \ln k$ plane derived from the WMAPII+SDSS chains with no $N$ constraint, showing $\Omega_m = \Omega_b + \Omega_{CDM}$, the late time variable which is most correlated with this eigenvector. This gives an indication of why the SDSS data does not help to break this degeneracy more strongly, despite the greater leverage on $k$-coverage one gets on adding it.} \label{fig:3param}
}

\FIGURE[!tb]{ 
\includegraphics[width=2.8in]{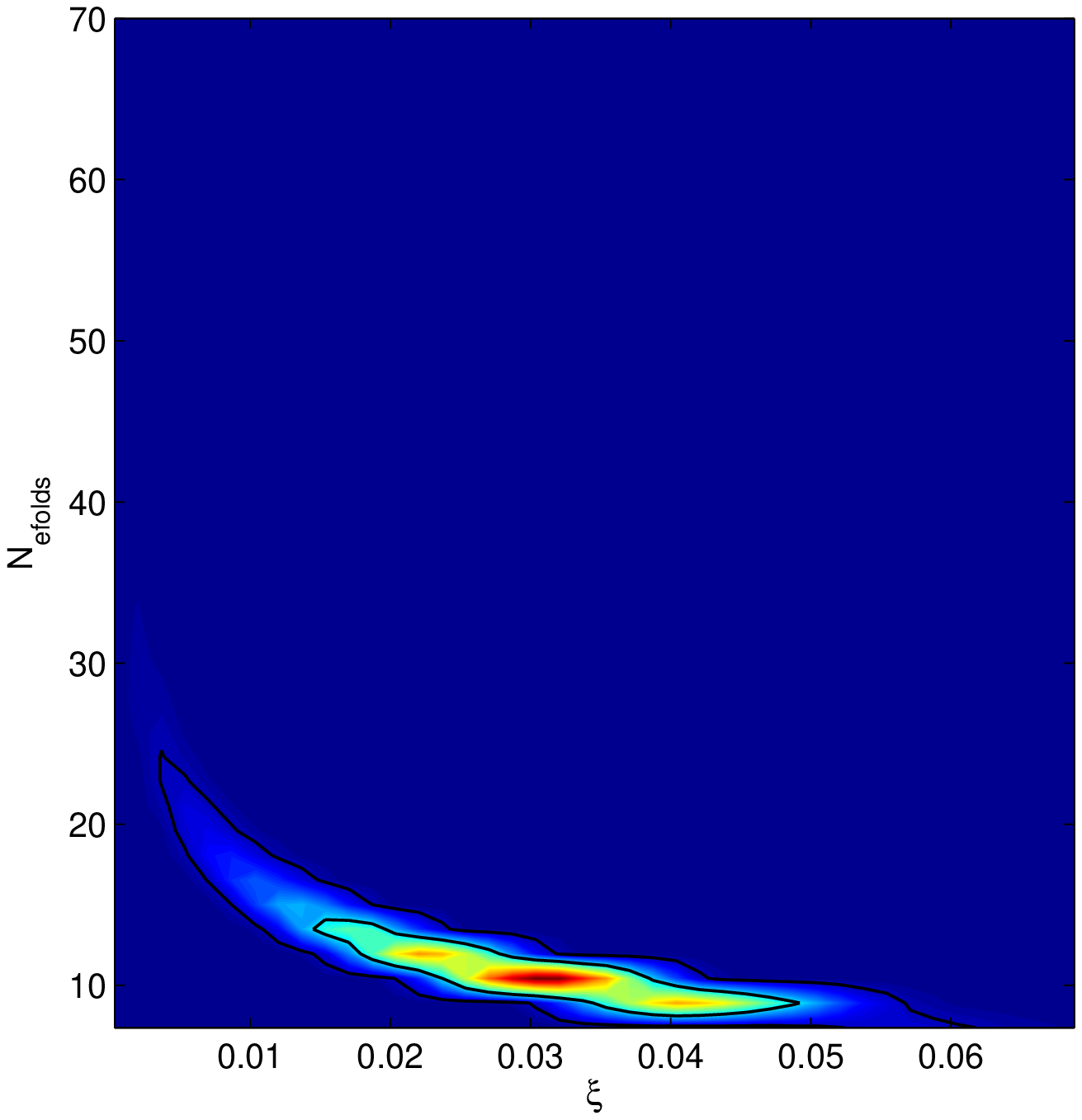}
\includegraphics[width=2.8in]{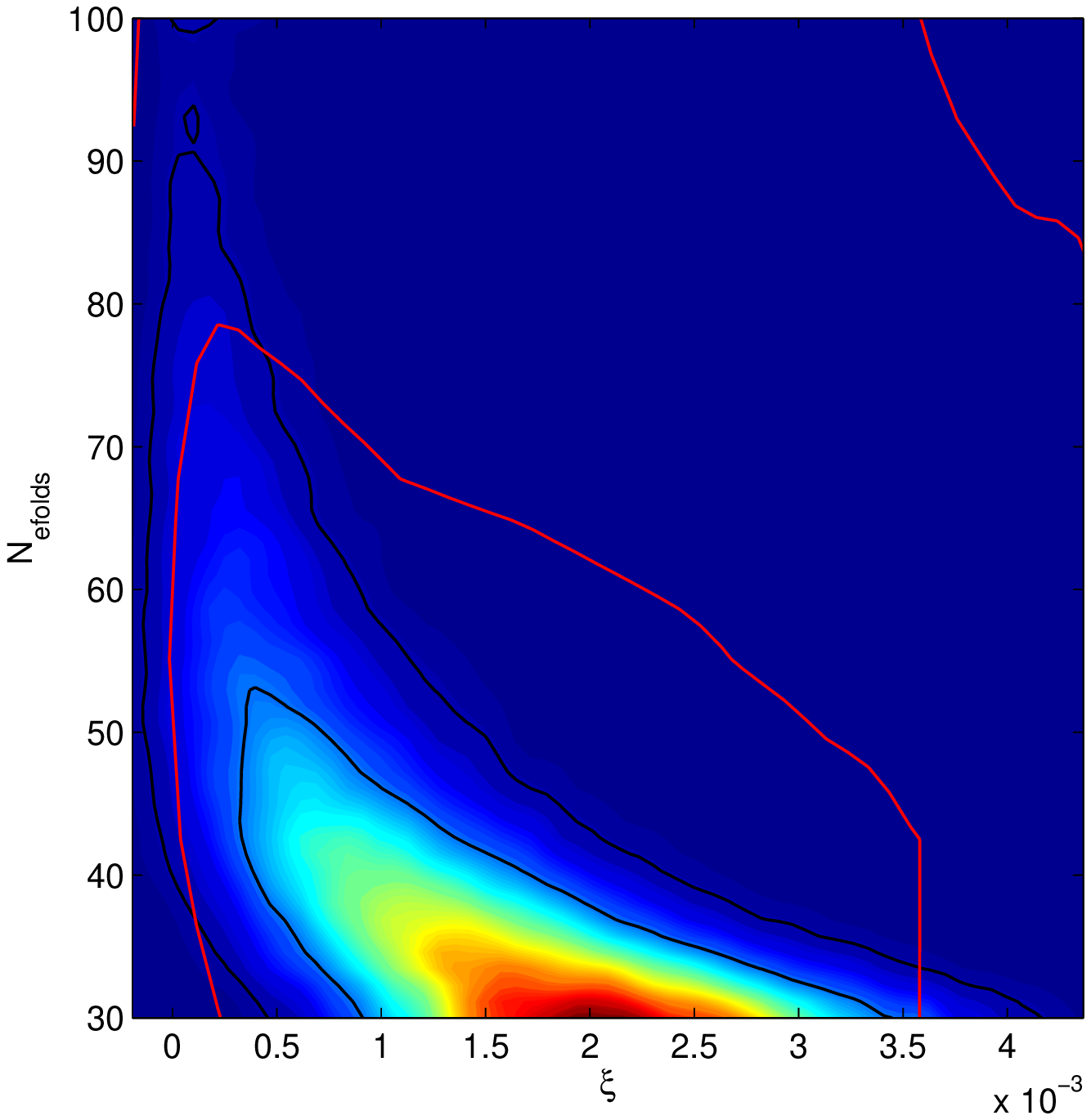}
\caption{The joint 68\% and 95\% confidence levels on the number of e-folds after the scale $k_0=0.002$ Mpc$^{-1}$ leaves the horizon till the end of inflation vs $\xi$ for WMAPII+SDSS (black) and WMAPII (red); no constraint on $N$ (left), $N>30$ (right). Only models from each set of chains where inflation explicitly ends are included.} \label{fig:Nbound}
}
 
\FIGURE[!tb]{ 
\includegraphics[width=2.5in]{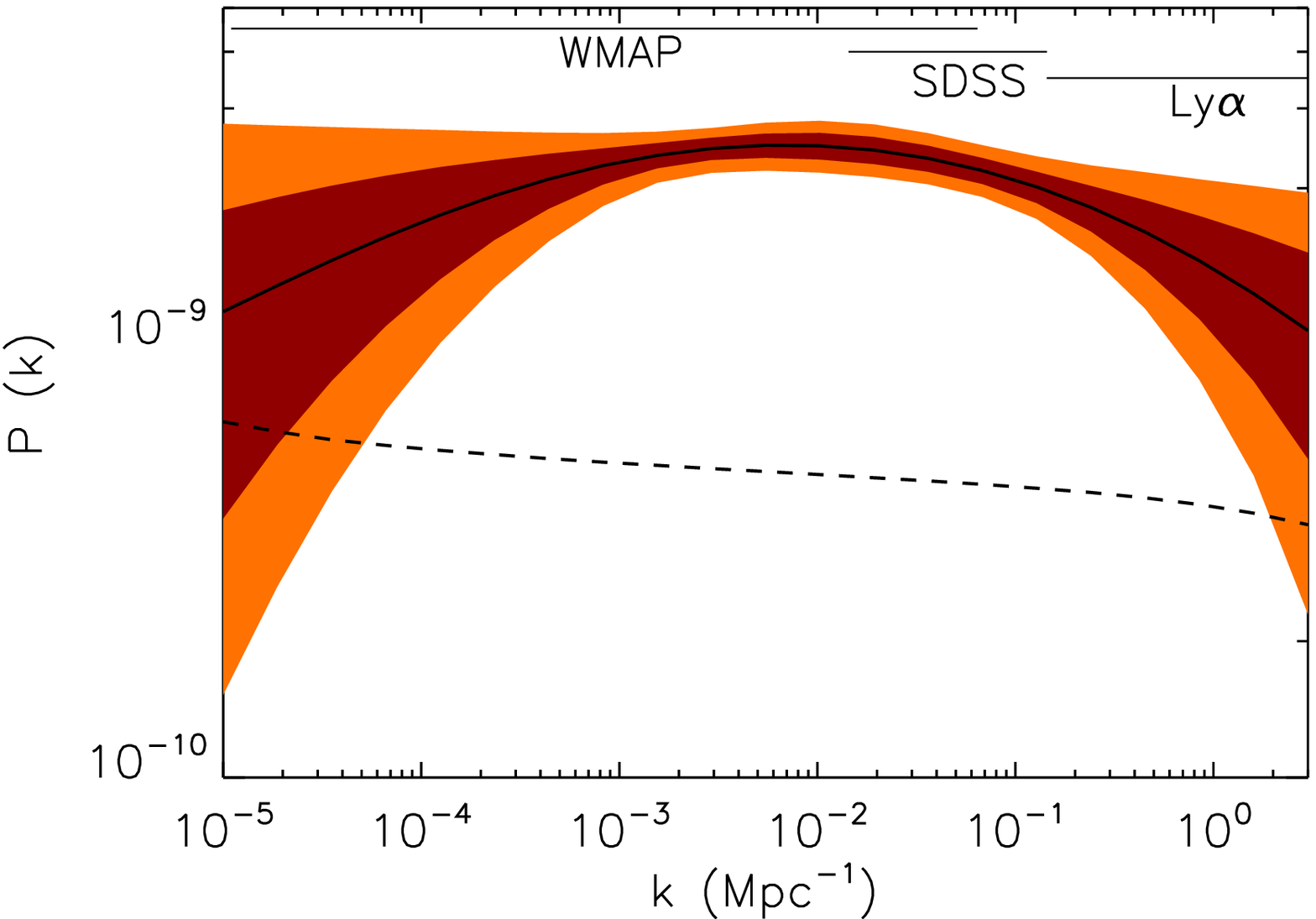} \\
\includegraphics[width=2.5in]{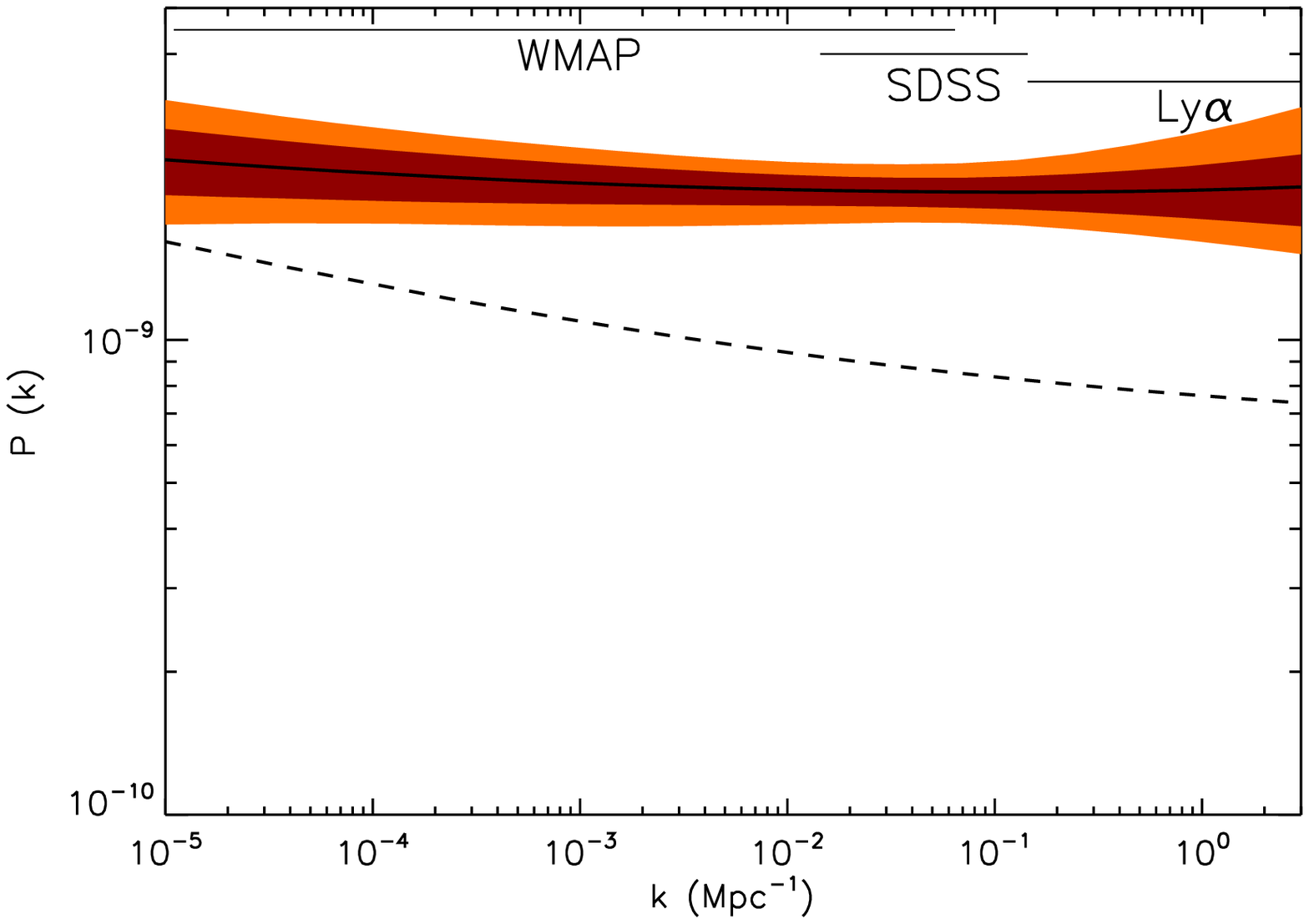} \hfill
\includegraphics[width=2.5in]{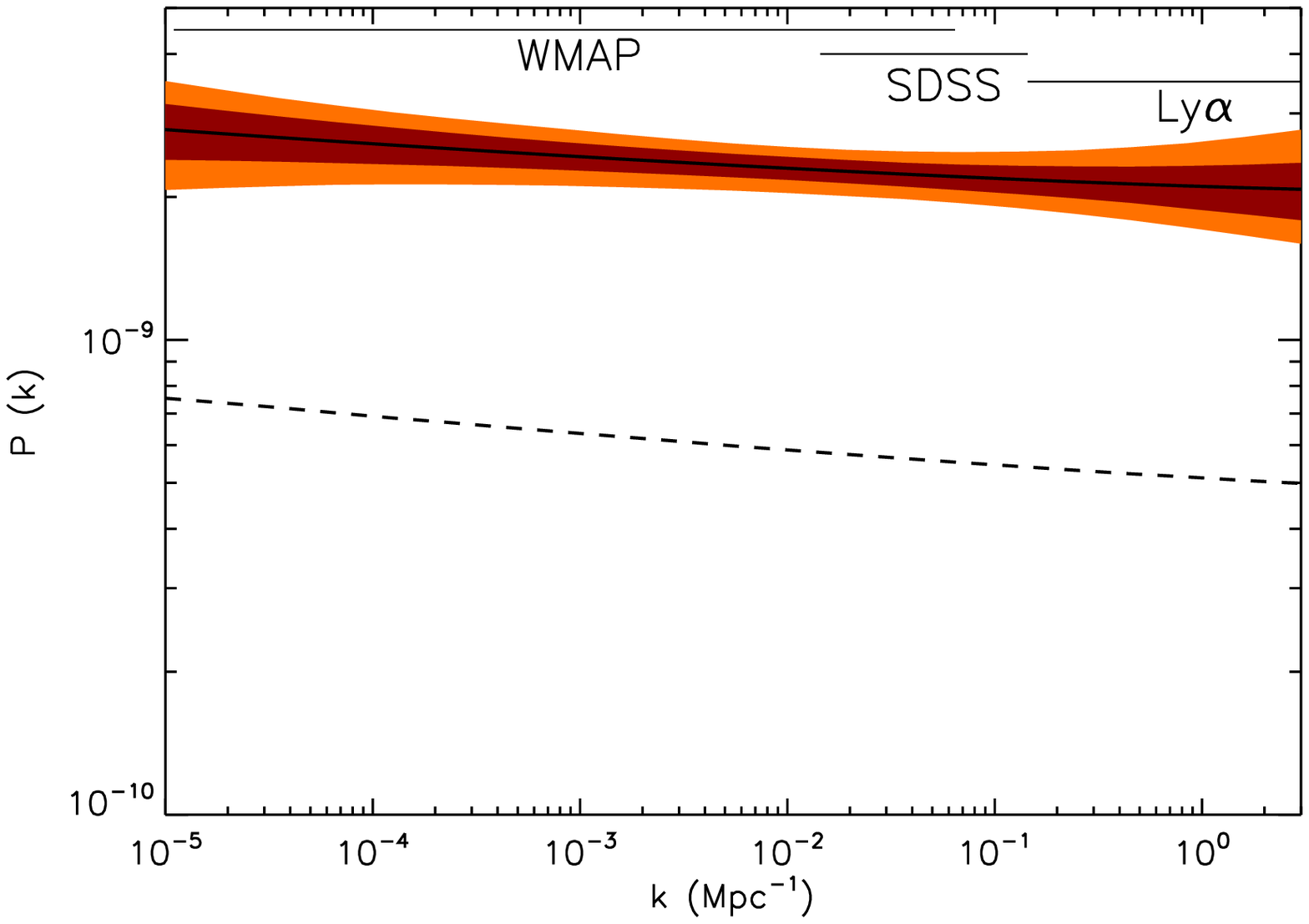}
\caption{Reconstructed 68\% (shaded dark) and 95\% (shaded light) constraints on the power spectrum of the primordial curvature perturbation as a function of $k$, shown for WMAPII+SDSS with no $N$ constraint (top), WMAPII with an $N>30$ prior (bottom left), and WMAPII+SDSS with an $N>30$ prior (bottom right). The solid line denotes the mean. The scales spanned by WMAP, SDSS and Lyman $\alpha$ data (not used in this analysis) are shown. The results with no $N$ constraint show a tendency of the power spectrum to run from blue at large scales to red at small scales, which disappears when the prior on $N$ is added. The dashed line shows the 95\% CL upper limit on the power spectrum of the primordial tensor perturbation.} \label{fig:pk}
}

\section{Interpretation and Analysis}

While slow roll reconstruction hinges on a very simple idea -- inserting the slow roll parameters directly into the cosmological parameter estimation process -- interpreting the results calls for a good deal of sophistication.   Without a constraint on the number of e-folds,  we see an apparent  preference for a large, positive value of $\xi$, which translates into  substantial and negative values of $\frac{dn_s}{d\ln{k}}$.    Recalling equation \ref{eq:pkempirical}, we see that if $\frac{dn_s}{d\ln{k}} \sim -0.05$ (near the center of the distribution derived from the combination   WMAPII+SDSS), the running term dominates this expression as we move away from $k_0$. As is well known, conventional cosmological observations probe the primordial spectrum over a range of around three orders of magnitude in scale. This may stretch to five orders in the future as we better probe the Lyman-$\alpha$ forest or obtain a power spectrum from observations of high redshift 21cm emission.  With $|\frac{dn_s}{d\ln{k}}| \sim 0.05$,  the second term in the exponent of equation \ref{eq:pkempirical} will dominate over the first for $k$ values that fall well inside the cosmological ``window'', and will dramatically modify the power relative to that seen at the pivot, $k_0$. For example, with $n_s\approx1$, $\ln{(k/k_0)} \approx 10$ and $\frac{dn_s}{d\ln{k}} \approx -0.05$,  the power spectrum at $k=e^{10} k_0$ is an order of magnitude smaller than at $k_0$.  However, if we have a spectrum that is strongly deficient in short scale power, the other variables in our parameter set will move  to offset this deficit.  As we can see from  Figure \ref{fig:lateparams}, this is exactly what happens here, as   $\Omega_m$ has increased relative to its concordance value, where $\Omega_\Lambda$ and $H_0$ have moved downwards.  The former is simply a result of our assumption of a flat universe, so that $\Omega_{\mbox{\tiny Total}} =1$, and the increased $\Omega_m$  compensates for the reduced power in the perturbation spectrum at short scales.    

Imposing additional constraints on $\Omega_m$ or $\Omega_\Lambda$ would put pressure on the models that produce large $\Omega_m$ and small $\Omega_\Lambda$.  In Figure~\ref{fig:3param} we show the distribution of $\{n_s,dn_s/d\ln{k}\}$ for Set 1 (WMAPII+SDSS), with the points color-coded by $\Omega_m$. This plot illustrates the correlation between an extreme running and an anomalously large $\Omega_m$.  Consequently,  a strong upper bound on $\Omega_m$ from observations  independent of the CMB and LSS data we employed in our MCMC analysis would exclude the models with the most extreme running. Consequently,  the long tail of the allowed region in the $\{n_s,dn_s/d\ln{k}\}$ plane for Set 1 can be understood as a marginalization artefact. 

In models with large $\xi$,   $\eta$ and $\epsilon$ are typically strongly scale-dependent, thanks to feedback from the $\xi$ term in the flow equations. Once $\epsilon$ is equal to unity, inflation will come to an end. As we have explained, imposing   $N>30$ leads to a dramatic change in the permitted region of parameter space,   consistent with our qualitative analysis in \cite{Easther:2006tv}.     In Figure~\ref{fig:Nbound}, we show the remaining number of e-folds as a function $\xi$, obtained by post-processing our chains.  Note that only models in which inflation explicitly ends are retained from the chains in order to make these plots. Models which are drawn to a ``late-time attractor''  and inflate forever (thus needing an orthogonal mechanism to end inflation) cannot be assigned an unambiguous number of remaining e-folds after the fiducial scale leaves the horizon. For this reason they are excluded from the figure (and we have checked that the Markov chains cut in this way still retain their convergence properties). We immediately see that Set 1 -- chains which ran without any constraint on $N$ -- has a distribution in the $\{\xi,N\}$ plane that {\em peaks\/} near $N=10$, and excludes $N=30$ at more than $2\sigma$.   On the other hand, the corresponding distribution for chains with $N>30$ is peaked  around $N=30$, but $N=60$ is  no longer excluded.

While we have dubbed the methodology employed here ``slow roll reconstruction'', we are not in a position to present an unambiguous form for the inflationary potential.  Without a non-zero lower bound on $\epsilon$ the height of the potential is essentially undetermined.  Secondly, as we noted above,  our chains prefer a positive value of $\xi$, but we emphasize that this is still well short of a convincing detection of a running, nor have we performed a Bayesian analysis (e.g., \cite{Parkinson:2006ku, Bridges:2006zm}) to test the significance of the inclusion of the $\xi$ parameter.  On the other hand, the underlying power spectrum is uniquely determined by any given choice of $\{A_s, \epsilon,\eta,\xi\}$, and in Figure~\ref{fig:pk}  we plot the permitted range of $P(k)$ for the three sets of chains.   In particular, without the $N>30$ prior we see the preference for a scale dependent spectral index in the WMAP+SDSS dataset. Conversely, it is easy to understand why including the Lyman-$\alpha$ dataset has such a dramatic impact on the allowable running, since this samples the power spectrum at very short scales, and thus has substantial ``leverage''.

Figures \ref{fig:Nbound} and \ref{fig:pk} show that large values of $\xi$ have a strong tendency to make inflation end after an insufficient number of e-folds to satisfy cosmological requirements. These figures imply that in order to satisfy the requirement that $N>30$, the allowed range of $\xi$ from the data becomes dramatically restricted, leading to a small allowed range on $\frac{dn_s}{d\ln{k}} \sim 0$.

While we specify the initial values of the slow roll parameters at a particular scale, we are able to post-process our chains to produce plots for the slow roll and spectral parameters at any arbitrary pivot (of course ensuring that the convergence criteria for the post-processed chains remain satisfied).  This is of particular importance when  deal with a running index, since $n_s$ is an explicit function of $k$, and the bounds on $n_s$ thus depend explicitly on $k_0$.  In our chains we took $k_0 = 0.002 \mbox{ Mpc}^{-1}$, but we now consider how the bounds on the parameter space vary as we translate the pivot.

\FIGURE[!tb]{ 
 \includegraphics[width=2.5in]{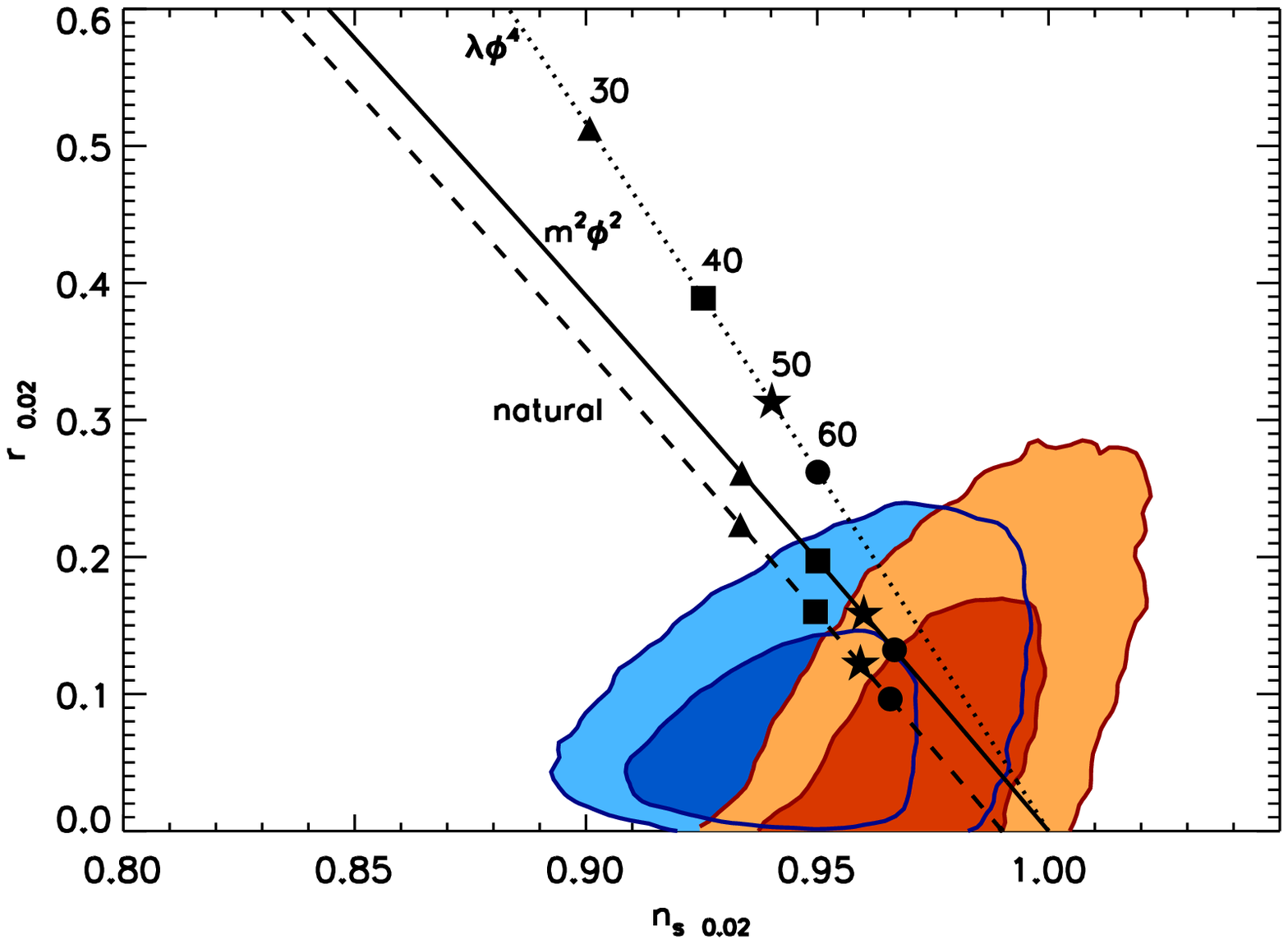}
\includegraphics[width=2.5in]{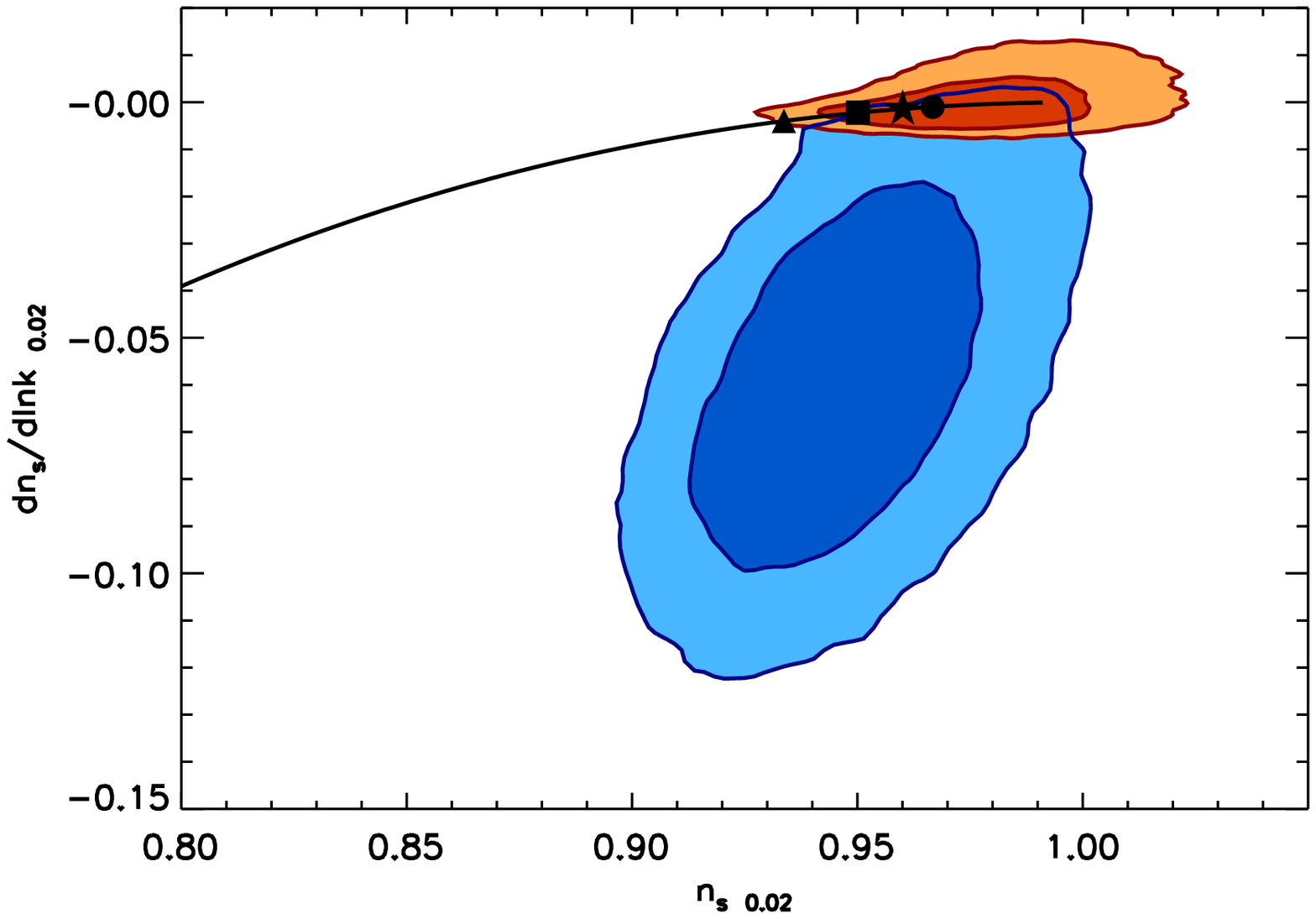}  \\
\includegraphics[width=2.5in]{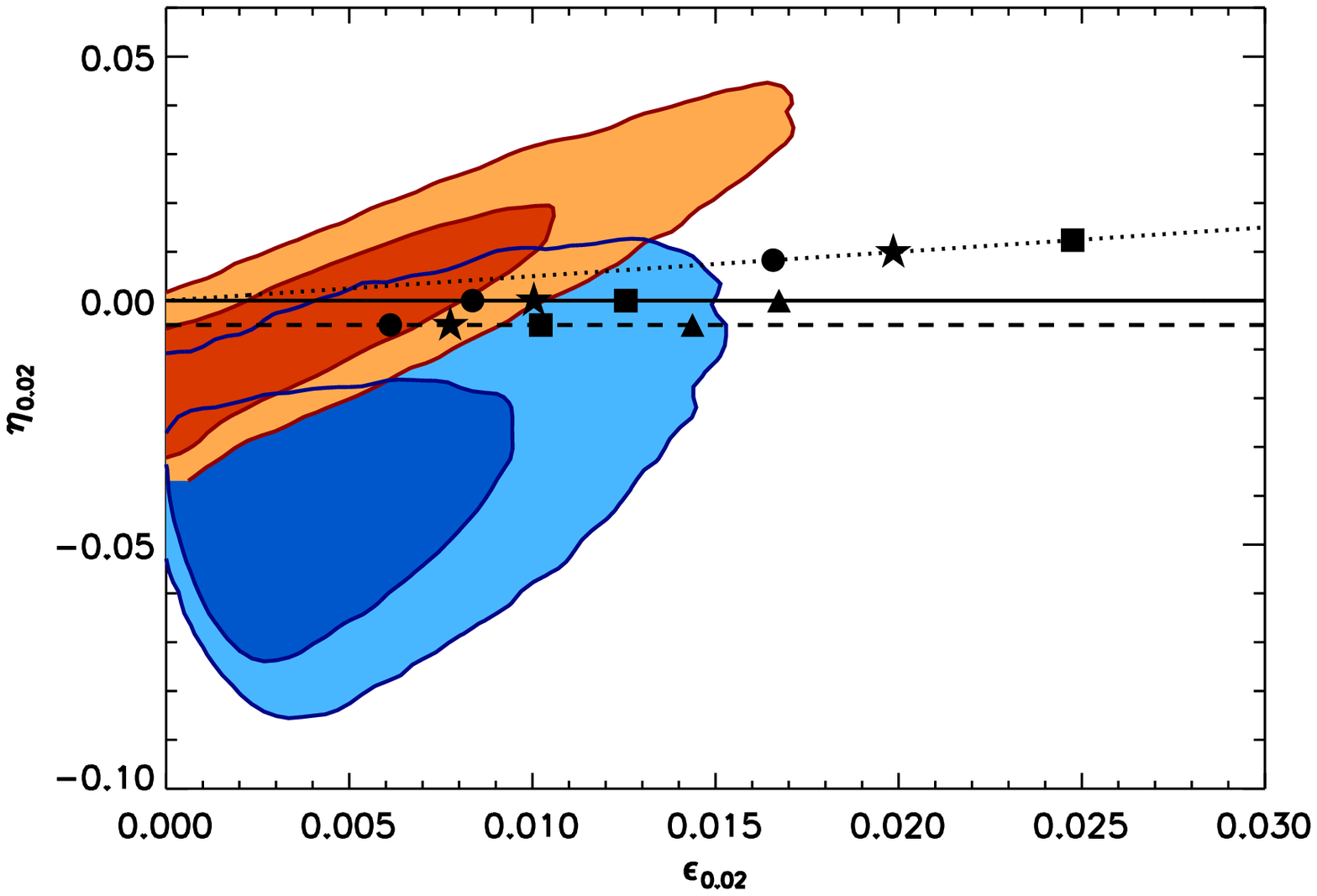} 
\includegraphics[width=2.5in]{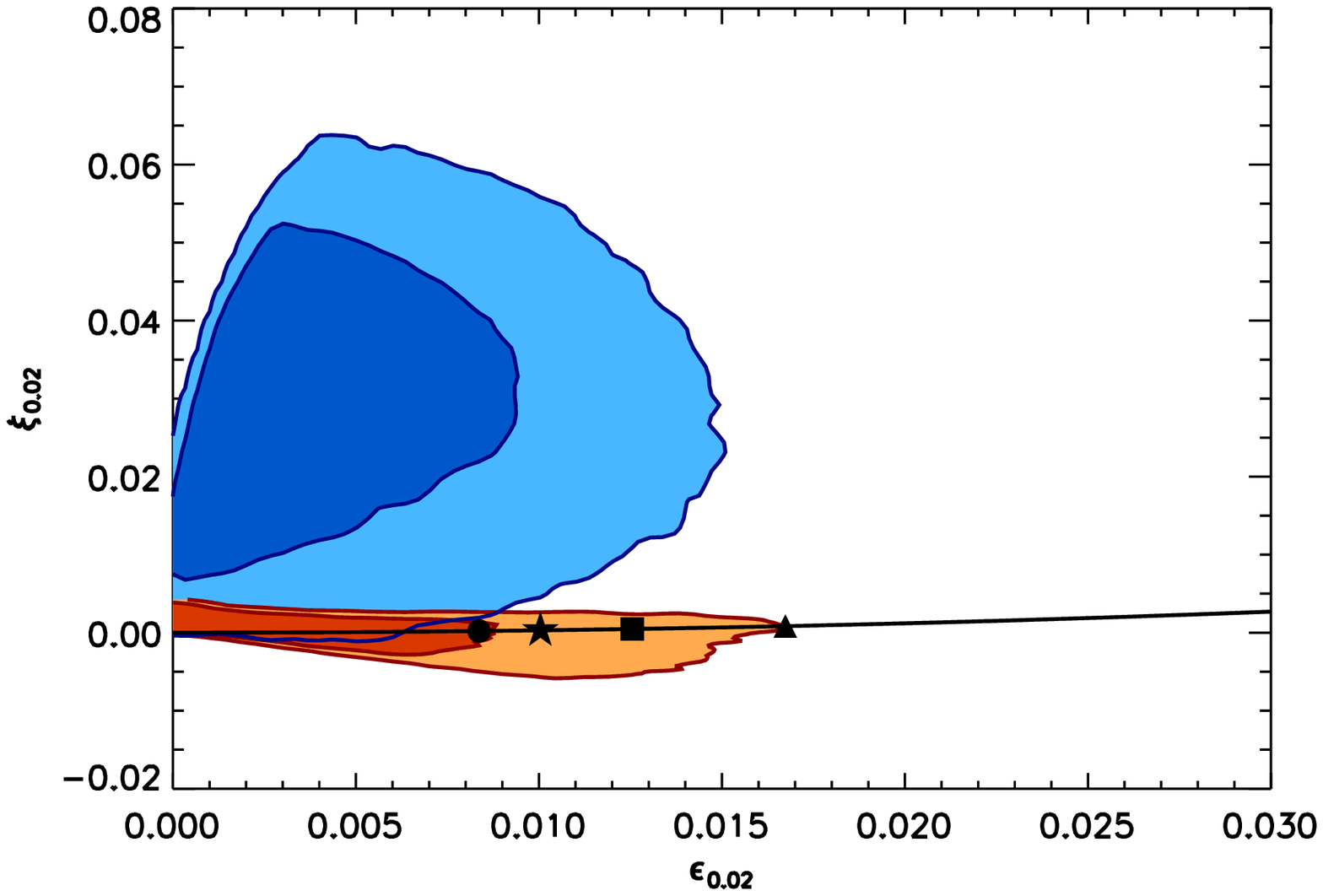} 
\caption{We present bounds on  the slow roll variables and spectral parameters, with the pivot moved to $k_0 = 0.02 \mbox{ Mpc}^{-1}$. 
The constraints are drawn from our WMAPII+SDSS chains,The blue regions are derived from Set 1 (no prior on $N$), and the brown from Set 3 ($N>30$).  We show 68\% and 95\% confidence intervals. 
We superimpose the ``trajectories'' for three generic  slow roll models, $\lambda \phi^4$ (dotted), $m^2 \phi^2$ (solid), and a representative natural inflation model (dashed), showing how the predictions of these models depend on $N$: $N=30$ (triangle), $N=40$ (square), $N=50$ (star), $N=60$ (circle). The figures containing running and $\xi$ only show the $m^2 \phi^2$ model, as the other models make similar predictions for the running.} \label{fig:newpivot}}

It is apparent from Figure \ref{fig:newpivot} that modifying $k_0$ significantly changes the constraints on the slow roll parameters and $n_s$ and $\frac{dn_s}{d\ln{k}}$.   Here we have chosen a pivot near the ``waist'' of the $P(k)$ distribution  in Figure~\ref{fig:pk}, or $k_0 = 0.02 \mbox{ Mpc}^{-1}$.  For the case where no $N$ prior was imposed, we now find that there is a considerable shift to the red in $n_s$ but the bounds on $\xi$ and $\frac{dn_s}{d\ln{k}}$ are not that much tighter than with $k_0 = 0.002 \mbox{ Mpc}^{-1}$. However, the $n_s$---$\frac{dn_s}{d\ln{k}}$ degeneracy has rotated somewhat, encapsulating both the tendency seen in the top panel of Figure \ref{fig:pk} for the power spectrum to run from blue on large scales to red on small scales, and the fact that specifying the constraints at the ``waist'' decorrelate these two parameters as much as possible. When we impose the $N$ prior, the constraints do not change significantly when we translate the pivot -- in this case  the running was already tightly constrained and the pivot dependence of $n_s$  is correspondingly  small.    We can compare these plots directly to specific inflationary models, and we have superimposed the trajectories for several  canonical models on the plots in Figure \ref{fig:newpivot}. A representative natural inflation \cite{Freese:1990rb} model \footnote{Natural inflation contains a free parameter $f$ that appears in the denominator of the cosine potential. Bounds derived from the WMAPI dataset on natural inflation are discussed in \cite{Freese:2004un}. } and $m^2 \phi^2$ inflation lie inside the 95\% confidence level with and without the $N$ prior for the combination of WMAPII+SDSS. Conversely, $\lambda \phi^4$ lies outside this confidence level for both cases. We do not show results for power-law or hybrid inflation. In the former case, $n_s$ and the slow roll parameters do not evolve with time, and so any given model would be represented by a single point on these plots, rather than a trajectory.  Likewise, in hybrid models the end of inflation is not specified by the value of the slow roll parameters, so we cannot unambiguously plot their trajectories  as a function of $N$.

Looking at Figure \ref{fig:newpivot} we are again reminded that interpreting MCMC estimates of cosmological parameters is a complicated task. At present, the publicly available Lyman-$\alpha$ likelihood codes are not compatible with the Hubble slow roll formalism used here. Since these data probe the fundamental spectrum at small scales, incorporating it into our analysis would greatly enhance our ability to constrain the spectrum -- particularly for models which are strongly scale dependent.   Likewise, we expect that the release of high quality CMB spectra for $\ell > 1000$ will greatly enhance the ability of slow roll reconstruction to constrain the inflationary parameter space. 

A further lesson to be drawn from Figure  \ref{fig:newpivot}  is that the data is now approaching the point where we must directly address the $N$ dependence of the parameters in simple inflationary models.  In \cite{Kinney:2005in}, Kinney and Riotto show that our ignorance of reheating physics means that we cannot make a unique mapping between the remaining number of e-folds $N$ and a given comoving scale $k$.  Looking at the trajectories we plot, we see that the $N$ dependence of the slow roll and spectral parameters over a ten e-fold range is only a few times smaller than  the widths of the parameter distributions  we recover from our MCMC analyses. The quality of cosmological data is thus reaching the point where this degeneracy will need to be included explicitly in any experimental tests of specific inflationary models -- especially those where the field point evolves comparatively rapidly.

\section{Discussion}

In this paper we have used slow roll reconstruction to  extract bounds on the inflationary potential from observational data. These bounds make strong use of an inflationary prior, and thus apply to cosmological models where the density perturbation spectrum is laid down by  single, slow rolling, minimally coupled inflaton field.   Our results align with the broad picture supplied by previous analyses of the WMAPII dataset: there is noticeable but by no means compelling evidence for a significant scale dependence (running) in the power spectrum when one considers WMAPII and SDSS on their own. However, adding a prior constraint that rules out models with an unacceptably small number of e-folds significantly tightens the bounds on any possible running.  

The combination of WMAPII and SDSS on their own are compatible with a broad class of perturbation spectra, but this range shrinks substantially when further constraints are added to the estimation process.  In particular, we consider inflationary models where at least 30 e-folds of inflation occur after CMB scales have left their horizon.  Similar results are seen in conventional analyses of the spectrum when Lyman-$\alpha$ forest data is included \cite{Seljak:2006bg}.   We further analyze the impact on the choice of pivot on the quoted ranges for the spectral parameters.  Slow roll reconstruction provides an estimate for the inflationary potential over its entire range, thanks to the closed nature of the truncated slow roll hierarchy.  Consequently, we actually have access to the underlying power spectrum and inflationary potential, although these quantities are fully specified by the truncated slow roll hierarchy -- which here consists of three independent parameters. However, we can use the flow equations to post-process our chains to give the constraints on the standard spectral parameters at an arbitrary pivot. We make use of this feature, and study the change in constraints on the inflationary parameter space when changing $k_0$ from  $0.002 \mbox{ Mpc}^{-1}$ to $0.02 \mbox{ Mpc}^{-1}$. Without a prior on the number of e-folds $N$, the constraints on $n_s$ and $\frac{dn_s}{d\ln{k}}$ are found to be significantly dependent on the pivot scale. The large running seen without this prior cannot be removed by shifting the pivot -- changing $k_0$ shifts $n_s$ well to the red, so the hints of an overall scale dependence cannot be expunged simply by changing $k_0$.  

One lesson from this analysis is that pivot-dependent estimates of power-law observables, while valid, need to be interpreted with great care. For example, in the case of the constraints presented at $k=$ 0.02 Mpc$^{-1}$ for WMAPII+SDSS without any prior on $N$, the fact that this pivot scale is an inflection point ``hides'' a tendency to run from blue on large scales to red on small scales as you move away from this pivot. That is, when there is a strong running, any slice in $k$ of the primordial power spectra can look quite different in $n_s$, $r$, $\frac{d n_s}{d\ln{k}}$ space. In our view, it is a better aid to interpretation to present constraints over the entire $k$ range implied by one's parameterization, as in Figure \ref{fig:pk} -- in principle,  this is a simple thing to do in the MCMC methodology for {\sl any} parameterization of $P(k)$ -- and then compare with specific inflationary models over this entire range of scales. What does matter is whether the constraints at a wide range of $k$ either encompass or exclude the prediction from that specific potential. For absolute completeness, the ``band'' of constraints from data on $P(k)$ should be compared with a ``band'' for the model under consideration, encompassing the theoretical uncertainty in both the model parameters and the reheating process  \cite{Kinney:2005in}.

Where they are directly comparable, our results broadly agree with  those of Finelli {\em et al.\/} \cite{Finelli:2006fi}.    However,  the authors of \cite{Finelli:2006fi} use a different formulation of the slow roll expansion from the one adopted here, which allows their $\epsilon_3$ to be of order unity, and they find they must pay careful attention to their prior for this parameter.\footnote{In the formalism of \cite{Finelli:2006fi}, $\epsilon_3$ only appears with a small factor, so while this parameter need not be small, its overall contribution is still typically sub-dominant.} Conversely, our ``third'' parameter, $\xi$ needs no such special handling, and we simply take all our slow roll parameters to lie within $[-1,1]$, which is far larger than the range allowed by the data. Secondly, our running ($\alpha$) is effectively scale-dependent, since $\xi$ is a function of scale, but  \cite{Finelli:2006fi} specify their running at a fixed pivot, and it is not a function of scale.  Finally, by using slow roll reconstruction we are able to include constraints on the remaining number of e-folds $N$, a possibility which is not considered in the analysis of  \cite{Finelli:2006fi}.    The tension between the running and the number of e-folds confirms our theoretical analysis in \cite{Easther:2006tv}.  

Algebraically, the running index is dominated by $\xi$, since $\epsilon$ and $\eta$ are both required by the data to be relatively small, and these terms only contribute quadratically to $\frac{dn_s}{d\ln{k}}$. When this parameter becomes large and positive the flow equations ensure that $\epsilon$ and $\eta$ will grow quickly. In this case $\epsilon$ becomes equal to unity in $\sim 15$ e-folds, and inflation terminates far too quickly for it to solve the usual cosmological problems.  In this analysis we have post-processed our chains to obtain likelihood contours for $N$ in the $(\xi,N)$ plane, and we see that $N$ does in fact peak around 10 e-folds. Conversely, if we impose a prior on $N$, the permitted range of $\frac{dn_s}{d\ln{k}}$ shrinks considerably.  

At large values, $\xi$ has a long degeneracy because of the lack of constraining power in the current dataset. This is possibly a marginalization artifact, as we found that a large running is correlated with values of $\Omega_m$ beyond the range seen in the standard concordance cosmology. If ``simple'' models of inflation are a good description of the universe, the rising quality of astrophysical data will eventually break this degeneracy, and the running will fall somewhere within the range found when the $N>30$ prior is applied here. Conversely if the data contracts around the large negative median value of the running, we would learn that inflation is non-minimal in some way, as discussed in \cite{Easther:2006tv}. In either case, slow roll reconstruction will be able to put tighter constraints on $N$ as the constraints on $\xi$ tighten.
 
In a subsequent paper we plan to assess the ability slow roll reconstruction to put bounds on the inflationary parameter space with the data from different proposed and projected experiments.  In addition, we will need to include $\lambda_3$, the fourth slow roll parameter, in these calculations, to be sure than any conclusions we reach are not a function of a premature truncation of the slow roll hierarchy. On the other hand, if this latter analysis shows that $\lambda_3$ is necessary to describe the data four independent parameters would be needed to describe the inflationary potential, which may be an indication that inflation is non-minimal.  

Finally we stress that while we find evidence that a significant breaking of scale-invariance in the primordial power spectrum is consistent with the 3-year WMAP dataset in the presence of an inflationary prior, there is no compelling evidence for a running spectral index. 
However, the analysis here gives us cause for optimism that this question can be settled in the near future, and that slow roll reconstruction provides an elegant and powerful framework for analyzing the cosmological constraints  on slow roll inflation.

\section*{Acknowledgments } 
 This work has made use of the {\sc LAMBDA} archive at {\tt http://lambda.gsfc.nasa.gov}.  RE is supported in part by the United States Department of Energy, grant DE-FG02-92ER-40704.  HVP is supported by NASA through Hubble Fellowship grant \#HF-01177.01-A awarded by the Space Telescope Science Institute, which is operated by the Association of Universities for Research in Astronomy, Inc., for NASA, under contract NAS 5-26555.  HVP wishes to thank the organizers and participants at the ``Inflation+25"  conference in Paris and the Benasque Cosmology Workshop in Summer 2006 for stimulating discussions which have helped to improve this paper, and Andrew Liddle for useful conversations. She acknowledges the hospitality of the IoA in Cambridge where part of this work was carried out. We thank Antony Lewis for prompt responses to questions about the GetDist parameter estimation package. We thank David Spergel and Lyman Page for valuable comments on an earlier draft.

\end{document}